\documentclass[aps,pra,reprint,showpacs,amsfonts,amsmath,floatfix,superscriptaddress,footinbib]{revtex4-2}

\usepackage[svgnames]{xcolor}
\usepackage{glossaries}

\usepackage[colorlinks=true,        
            allcolors = black,  
            citecolor=DarkBlue, 
           linkcolor=DarkBlue,
            urlcolor=DarkBlue
        ]{hyperref}  
\usepackage{amsmath, amsfonts, amssymb, amsthm, bbm}

\usepackage{dsfont}
\usepackage{bm} 
\usepackage{mathtools}
\usepackage{physics}
\usepackage{graphicx}

\usepackage{dcolumn}

\usepackage{tikz}
\usepackage{float}
\usepackage{soul}
\usepackage{xcolor}

\usepackage{array,multirow}

\usepackage{bbold}
\usepackage{comment} 

\newcommand{\pauli}[2]{\hat{\sigma}_{#1}^{#2}}
\newcommand{\transfer}[2]{\eta_{{#1},{#2}}}
\newcommand{\gammapar}{\bm{\gamma}}
\newcommand{\betapar}{\bm{\beta}}
\newcommand{\mixerH}{\hat{H}_M}
\newcommand{\costH}{\hat{H}_C}
\newcommand{\donor}{D}
\newcommand{\recep}{R}

\newacronym{vqa}{VQA}{variational quantum algorithm}
\newacronym{qaoa}{QAOA}{quantum approximate optimization algorithm}
\newacronym{nisq}{NISQ}{noisy intermediate-size quantum}
\newacronym{cop}{COP}{combinatorial optimization problem}
\newacronym{ows}{OWS}{odd weight sum}
\newacronym{ews}{EWS}{even weight sum}
\newacronym{rqaoa}{RQAOA}{recursive QAOA}
\newacronym{ma-qaoa}{ma-QAOA}{multi-angle QAOA}
\newacronym{vqe}{VQE}{variational quantum eigensolver}
\newacronym{varqite}{VarQITE}{variational quantum imaginary time evolution}
\newacronym{hva}{HVA}{Hamiltonian variational ansatz}
\newacronym{tfim}{TFIM}{transverse-field Ising model}

\begin{document}

\title{Symmetry-informed transferability of optimal parameters in the Quantum Approximate Optimization Algorithm}

\author{Isak Lyngfelt}%
\email{isak.lyngfelt@chalmers.se}
\author{Laura García-Álvarez}
\affiliation{%
Department of Microtechnology and Nanoscience (MC2), Chalmers University of Technology, SE-412 96 G\"{o}teborg, Sweden
}

\begin{abstract}
One of the main limitations of variational quantum algorithms is the classical optimization of the highly dimensional non-convex variational parameter landscape. To simplify this optimization, we can reduce the search space using problem symmetries and typical optimal parameters as initial points if they concentrate.
In this article, we consider typical values of optimal parameters of the quantum approximate optimization algorithm for the MaxCut problem with $d$-regular tree subgraphs and reuse them in different graph instances. 
We prove symmetries in the optimization landscape of several kinds of weighted and unweighted graphs, which explains the existence of multiple sets of optimal parameters. However, we observe that not all optimal sets can be successfully transferred between problem instances.
We find specific transferable domains in the search space and show how to translate an arbitrary set of optimal parameters into the adequate domain using the studied symmetries.
Finally, we extend these results to general classical optimization problems described by Ising Hamiltonians, the Hamiltonian variational ansatz for relevant physical models, and the recursive and multi-angle quantum approximate optimization algorithms.

\end{abstract}
\maketitle

\section{Introduction}
\Glspl{vqa} are among the most popular and accessible quantum algorithms for solving classically hard problems, with a notable example being the \gls{qaoa}~\cite{farhiQuantumApproximateOptimization2014}, used to tackle \glspl{cop}. 
The \gls{qaoa} is a hybrid algorithm consisting of a quantum circuit with several layers of parameter-dependent operations applied iteratively, followed by a classical optimization of the parameters.
The quality of the solutions output by this algorithm depends on these classically optimized variational parameters and the depth of the quantum circuit. 
In theory, the solution improves as the circuit depth increases but, in practice, deep circuits introduce noise that affects the quantum state. However, shallow depth \gls{qaoa} is often inferior to state-of-the art classical algorithms~\cite{farhiQuantumApproximateOptimization2020, farhiQuantumApproximateOptimization2020a, bassoPerformanceLimitationsQAOA2022, anshuConcentrationBoundsQuantum2023, depalmaLimitationsVariationalQuantum2023}.
Therefore, we have to choose an algorithm depth that achieves an acceptable solution while maintaining quantum coherence.
Alternatively, one can modify the standard \gls{qaoa} to mitigate the effects of decoherence. Multiple alternative \gls{qaoa} inspired algorithms adapted to specific \glspl{cop} show better theoretical performance at lower depths, like the \gls{rqaoa}~\cite{bravyiObstaclesVariationalQuantum2020} or the \gls{ma-qaoa}~\cite{herrmanMultiangleQuantumApproximate2021}. Yet their success also depends on finding good variational parameters.

The number of variational parameters scales linearly with the depth of the circuit, and the number of random initialization points needed to find the global optima is exponential in the number of parameters~\cite{zhouQuantumApproximateOptimization2020}. 
All \gls{qaoa} variants suffer from this curse of dimensionality and, in general, the classical optimization routine of \gls{qaoa} is known to be an NP-hard problem~\cite{bittelTrainingVariationalQuantum2021}. Barren plateaus in the optimization landscape further increase the difficulty in finding the optimal parameters~\cite{cerezoVariationalQuantumAlgorithms2021,holmesConnectingAnsatzExpressibility2022, bittelTrainingVariationalQuantum2021, ragoneUnifiedTheoryBarren2023}, and it has been suggested that problems that are classically hard to simulate will always have these plateaus~\cite{cerezoDoesProvableAbsence2023}. As a result, efficient strategies for finding optimal parameters are vital in the viability of \gls{qaoa}~\cite{rajakumarTrainabilityBarriersLowDepth2024}.

Optimizing the variational parameters is an active field of research, and a multitude of techniques have been studied, including standard classical optimizers~\cite{fernandez-pendasStudyPerformanceClassical2022}, gradient-free optimizers~\cite{fernandez-pendasStudyPerformanceClassical2022,mccleanTheoryVariationalHybrid2016}, classical genetic algorithms~\cite{acamporaGeneticAlgorithmsClassical2023}, machine-learning assisted optimization~\cite{alamAcceleratingQuantumApproximate2020, wangAutomatedQuantumCircuit2023, deshpandeCapturingSymmetriesQuantum2022,Cheng2024}, and extrapolation techniques~\cite{zhouQuantumApproximateOptimization2020,leeDepthProgressiveInitializationStrategy2023}. Furthermore, we can benefit from prior knowledge of the problem, either by initializing a quantum state close to the solution~\cite{eggerWarmstartingQuantumOptimization2021, cainQAOAGetsStuck2023,tateGuaranteesWarmStartedQAOA2024}, finding a better initial ansatz by an efficient classical optimization over Clifford gates~\cite{Mu_oz_Arias_2024, ravi2023cafqaclassicalsimulationbootstrap} or bypassing the optimization by transferring pre-trained optimal parameters~\cite{brandaoFixedControlParameters2018,galdaSimilarityBasedParameterTransferability2023,streifTrainingQuantumApproximate2020,shaydulinParameterTransferQuantum2023, sureshbabuParameterSettingQuantum2024, montanez-barreraTransferLearningOptimal2024, fallaGraphRepresentationLearning2024, shaydulinEvidenceScalingAdvantage2024, lotshawEmpiricalPerformanceBounds2021, PhysRevA.106.L060401}. The optimization can also be simplified by employing fixed schemes such as Fourier-series extrapolation~\cite{zhouQuantumApproximateOptimization2020}, discrete adiabatic paths~\cite{zhouQuantumApproximateOptimization2020, leeDepthProgressiveInitializationStrategy2023}, or linear ramp schedules~\cite{montanez-barreraUniversalQAOAProtocol2024}.
While the main purpose of \gls{qaoa} is finding approximate solutions to \glspl{cop}, the sampling from the output quantum state is related to pseudo-Boltzmann distributions that are classically intractable~\cite{diez-valleQuantumApproximateOptimization2023, lotshawApproximateBoltzmannDistributions2023}. In these cases, the quality of the algorithm parameters affects the effective temperature of the probability distribution, with the optima leading to the lowest temperature. 

In some cases, the optimal parameters concentrate on typical values in parameter space. This behavior was initially observed for MaxCut on $3$-regular graphs~\cite{bassoQuantumApproximateOptimization2022}, i.e., graphs where every vertex is connected to 3 vertices, and later for MaxCut on other kinds of graphs~\cite{zhouQuantumApproximateOptimization2020, diez-valleQuantumApproximateOptimization2023,farhiQuantumApproximateOptimization2022, sureshbabuParameterSettingQuantum2024}. This phenomenon has been linked with the locality of \gls{qaoa} at shallow depths, where the cost Hamiltonian of a graph can be reformulated as a sum of local Hamiltonians of subgraphs~\cite{farhiQuantumApproximateOptimization2014, zhouQuantumApproximateOptimization2020}. The typical optimal parameters are those optimal to the local Hamiltonian that dominate the expectation value of the total cost Hamiltonian~\cite{brandaoFixedControlParameters2018, streifTrainingQuantumApproximate2020, lykovSamplingFrequencyThresholds2023, galdaSimilarityBasedParameterTransferability2023}. At algorithm depth $p=11$, numerical studies show that MaxCut on $3$-regular graphs with typical parameters has performance guarantees superior to the best classical algorithms~\cite{wurtzFixedangleConjecturesQuantum2021}, and a recent work by Ref.~\cite{wybo2024missingpuzzlepiecesperformance} suggests that the \gls{qaoa} with typical parameters already has an advantage at $p=8$.
Furthermore, it is possible to achieve near-optimal \gls{qaoa} performances by transferring the typical values from a donor problem instance to a receiver problem instance drawn from a different ensemble. These successful transfers have been numerically confirmed between regular graphs with either even or odd degrees~\cite{galdaSimilarityBasedParameterTransferability2023, galdaTransferabilityOptimalQAOA2021}, between the Sherrington-Kirkpatrick model and MaxCut on high-girth graphs~\cite{farhiQuantumApproximateOptimization2022, bassoQuantumApproximateOptimization2022}, between unweighted and weighted graphs~\cite{shaydulinParameterTransferQuantum2023,sureshbabuParameterSettingQuantum2024}, and between the MaxCut problem and other \glspl{cop}~\cite{montanez-barreraTransferLearningOptimal2024}. 
Additionally, the transferred parameters can be used as initial guesses for a warm-start local optimization of \gls{qaoa} parameters, significantly accelerating the classical loop~\cite{alamAcceleratingQuantumApproximate2020, zhouQuantumApproximateOptimization2020,shaydulinMultistartMethodsQuantum2019, fallaGraphRepresentationLearning2024}. 
Warm-start optimization is a promising heuristic method to mitigate the challenge of barren plateaus. Unlike other methods to avoid barren plateaus which often inadvertently imply that the problem at hand is efficiently solvable by classical computers, warm-start techniques are not believed to have this limitation~\cite{cerezoDoesProvableAbsence2023}.

The classical optimization can also be simplified by identifying symmetries in the parameter space and limiting the search space accordingly~\cite{zhouQuantumApproximateOptimization2020,shaydulinExploitingSymmetryReduces2021,lotshawEmpiricalPerformanceBounds2021}.
In principle, any optimal parameters related by symmetry operations output the same quantum state and, thus, perform equally when solving the classical problem. 
However, when transferring these parameters to other instances, not all sets of optimal parameters are suitable~\cite{galdaSimilarityBasedParameterTransferability2023, galdaTransferabilityOptimalQAOA2021}. 
In particular, one should consider domains in the search space containing optimal parameters for both the donor and the receiver instances. 
To this end, it is essential to study the symmetries in both the donor and receiver problem instances.

In this article, we identify symmetries in the optimization landscape of \gls{qaoa} for Ising Hamiltonian with certain integer coefficients previously not addressed in the literature. In addition, we study the performance of \gls{qaoa} for MaxCut using transferred parameters between problem instances for circuit depths $p=1$ and $p=2$. 
We define a function characterizing the transferability of parameters and provide an analytical closed-form expression for triangle-free regular graphs.  Such graphs are said to have a girth higher than 3, i.e., the shortest cycle length is longer than 3. We observe that, in general, these transferred parameters reach a high performance if they belong to certain domains in the search space. 
Our analysis extends to $d$-regular graphs with integer weights and unweighted random graphs from different models---graphs with a fixed number of vertices, and edges generated by different probability distributions. We show that when specific parameters are transferable, all parameters can be effectively transferred by applying translations based on the symmetries of the variational state's expectation value. This result appears to extend to higher algorithm depths, suggesting a scalable approach to parameter optimization. Furthermore, while our numerical study focuses on particular problem instances, our work provides tools to analyze the symmetries and transferability of parameters, which can help to identify successful initialization or optimization schemes in general \glspl{vqa}. Building on this idea, we extend our symmetry results to the \gls{hva}, a relevant quantum heuristic for physical models in the context of \glspl{vqe}, as well as the \gls{rqaoa} and the \gls{ma-qaoa}. For the latter two, we also discuss the potential of parameter transferability.

Different initialization or optimization schemes have unique strengths and weaknesses that make them more suitable for specific \glspl{vqa}. For hardware-efficient ansätze, a classically efficient search over Clifford-gate constructed ansätze has proven effective for finding good initializations~\cite{ravi2023cafqaclassicalsimulationbootstrap}. However, this approach is less suitable for structured ansätze that utilize problem-specific features, such as \gls{qaoa} or \gls{hva} circuits. Indeed, in \gls{qaoa} circuits, Clifford gates constrain the parameters to integer multiples of $\pi/4$, which are typically far from the optimal values~\cite{Mu_oz_Arias_2024}, unless one considers hardware-efficient \gls{qaoa} variants, like ADAPT-\gls{qaoa}~\cite{PhysRevResearch.4.033029}. Conversely, our results are not directly applicable to hardware-efficient ansatz circuits, as we rely on integer-valued coupling constants in the spin model to derive parameter space symmetries. These assumptions suggest that our findings will improve parameter training in structured quantum circuits for specific problems with integer-valued coupling constants in the Hamiltonians. 

The article is structured as follows. In Sec.~\ref{sec:SymmetriesQAOA}, we review the formulation of \gls{qaoa} and the \gls{cop} MaxCut, as well as the known symmetries of the variational parameters. We prove a generalization of these symmetries to include a wider family of graphs. Then, in Sec.~\ref{sec:transferability}, we analyze the transferability of optimal variational parameters and identify which optimal sets are suitable for reuse. In Sec.~\ref{sec:extension}, we generalize our symmetries to encompass more \glspl{vqa} and discuss the parameters' transferability in those cases. Lastly, we summarize our results in Sec.~\ref{sec:outlook}.

\section{Symmetries of QAOA for MaxCut}
\label{sec:SymmetriesQAOA}

We begin by reviewing the formulation of QAOA, an algorithm that can be easily adapted to tackle the optimization of classical functions, and introducing the notation we use throughout the manuscript.
In this work, we consider a fundamental NP-complete problem in graph theory, namely MaxCut, whose solution is also hard to approximate. This problem seeks to divide the vertices of an undirected graph into two sets to maximize the total weight of edges---or the number of edges in the case of unweighted graphs---crossing between these sets. That is, given an undirected graph $G=(V,E)$ with weights $w_{ij}$ on the edges $(i,j)\in E$ joining the vertices $V$, we seek to maximize
\begin{equation}
\label{eq:clasMaxCut}
    C(x)= \frac{1}{2}\sum_{(i,j)\in E} w_{ij}(1-x_i x_j),
\end{equation}
with $x\in \{-1,1\}^{n}$ the length-$n$ string to be optimized, and $n$ the order of the graph $|V|$. For a given problem instance, the quality of a feasible solution $x$ can be quantified by the approximation ratio,
\begin{equation}
\label{eq:approxratioclas}
    r(x)= \frac{C(x)-C_{\rm{min}}}{C_{\rm{max}}-C_{\rm{min}}},
\end{equation}
which is equal to one for optimal solutions and approaches zero as the quality decreases.
For MaxCut, the best classical algorithms guarantee an approximation ratio of at least 0.8785~\cite{goemansImprovedApproximationAlgorithms1995}.

The MaxCut problem can be encoded on a quantum computer by rewriting the cost function of Eq.~(\ref{eq:clasMaxCut}) in terms of quantum operators. We replace each variable $x_i$ by the Pauli operator $\pauli{z}{i}$ to obtain the cost Hamiltonian
\begin{equation}
\label{eq:HC}
    \costH = \frac{1}{2}\sum_{(i,j)\in E} w_{ij} \left(\mathbb{1}-\pauli{z}{i}\pauli{z}{j}\right),
\end{equation}
where the problem's solutions are related to the measurement outcomes $\pm 1$ of each qubit in the computational basis.
The \gls{qaoa} leverages the problem structure by encoding it in the cost unitary $\hat{U}_C(\gamma) = e^{-i\gamma\costH}$. The algorithm iteratively applies this cost unitary and a mixer unitary $\hat{U}_M(\beta) = e^{-i\beta\mixerH}$---with the mixer Hamiltonian $\mixerH=\sum_i\pauli{x}{i}$---to the symmetric superposition of computational basis states $\ket{+}^{\otimes n}$ to prepare the $n$-qubit variational state $\ket{\gammapar,\betapar}$. The state depends on the real-valued parameters $\gammapar=(\gamma_1, \dots, \gamma_p)$ and $\betapar=(\beta_1, \dots, \beta_p)$ assigned to the cost and mixing operations for the total $p$ iterations---depth $p$---and aims to optimize the expectation value of the cost Hamiltonian $\costH$,
\begin{equation}
\label{eq:Cp}
    C(\gammapar,\betapar) = \bra{\gammapar,\betapar}\costH\ket{\gammapar,\betapar}.
\end{equation} 
We can thus consider the approximation ratio over the distribution of \gls{qaoa} solution outputs as a quality measure,
\begin{equation}
\label{eq:approxratioq}
    r(\gammapar,\betapar)= \frac{C(\gammapar,\betapar)-C_{\rm{min}}}{C_{\rm{max}}-C_{\rm{min}}}.
\end{equation}

The cost landscape represents the relationship between the objective function of Eq.~(\ref{eq:Cp}) and the parameters $\gammapar$ and $\betapar$. The classical optimization of $\gammapar$ and $\betapar$ is generally challenging due to highly non-convex landscapes containing several local minima and maxima. The landscape symmetries help simplify the optimization process by restricting the search space.
In other words, these symmetries help divide the search space into domains containing optimal parameters that perform equally when finding good approximate solutions to a \gls{cop} or creating final states from which one can perform ``pseudo-thermal" sampling~\cite{lotshawApproximateBoltzmannDistributions2023, diez-valleQuantumApproximateOptimization2023}.

\begin{figure}
    \centering
    \includegraphics[width = 1\linewidth]{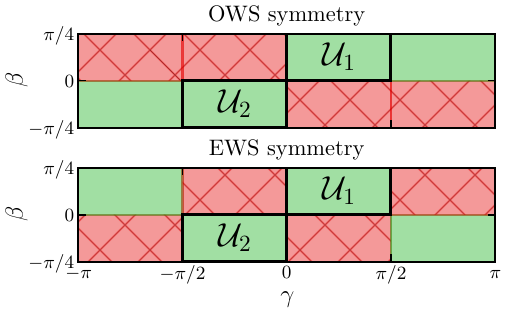}
    \caption{Symmetries for optimal \gls{qaoa} parameters $(\gamma,\beta)$ at depth $p=1$ for unweighted regular graphs with odd (top) and even (bottom) degrees, which follow OWS and EWS symmetries, respectively. Each solid green area contains one set of optimal parameters, while they do not appear in red-crossed areas. The domains $\mathcal{U}_1$ and $\mathcal{U}_2$ contain optimal parameters in both cases.}
    \label{fig:angles_symmetry}
\end{figure}

First, we notice that the cost and mixer Hamiltonians are real symmetric matrices in the computational basis and, therefore, \gls{qaoa} exhibits a ``time-reversal'' symmetry $(\gammapar,\betapar)\to (-\gammapar,-\betapar)$~\cite{zhouQuantumApproximateOptimization2020, lotshawEmpiricalPerformanceBounds2021}. This symmetry inverts the sign of all elements of the vectors simultaneously and is valid for any graph with real weights.
Second, we consider the standard mixer Hamiltonian $\mixerH = \sum_i \pauli{x}{i}$ and MaxCut problem instances with integer weights so that the cost Hamiltonian has integer eigenvalues and only two-qubit interactions. Therefore, we begin by restricting the parameter optimization to the domain $\mathcal{A}^p$, such that $\gammapar \in [-\pi,\pi)^{p}$ and $\betapar \in [-\pi/4,\pi/4)^{p}$---as shown in Fig.~\ref{fig:angles_symmetry} for $p=1$---since the \gls{qaoa} cost function for MaxCut remains unchanged when any $\gamma_i$ and $\beta_i$ parameters shift by integer multiples of $2\pi$ and $\pi/2$, respectively. 

In the following, we review additional symmetries of the optimal parameters within $\mathcal{A}=[-\pi,\pi)\times [-\pi/4,\pi/4)$, summarized in Table~\ref{tab:angle_symmetry_general}, and extend the known results for MaxCut unweighted graph instances~\cite{zhouQuantumApproximateOptimization2020,lotshawEmpiricalPerformanceBounds2021} to specific families of weighted graphs.

\begin{table}
\caption{\label{tab:angle_symmetry_general}Transformations of $(\gammapar,\betapar)\to(\gammapar',\betapar')$ such that $C(\gammapar,\betapar)=C(\gammapar',\betapar')$. The \gls{ews} symmetry refers to integer-weighted graphs where the sum of weights connected to each vertex is even, i.e., $\Sigma_jw_{jk}\in 2\mathbb{Z}, \, \forall k\in V$. Analogously, \gls{ows} symmetry refers to integer-weighted graphs where the sum of the weights connected to each vertex is odd, i.e., $\Sigma_jw_{jk}\in\mathbb{Z}\setminus2\mathbb{Z}, \,\forall k\in V$.}

\begin{ruledtabular}
\begin{tabular}{ll}
\multicolumn{1}{c}{\textbf{Graph}} & \multicolumn{1}{c}{\textbf{Symmetry}} \\ 
\gls{ews} & $(\gamma_1,...,\gamma_i,...,\gamma_p)\to(\gamma_1,...,\gamma_i\pm\pi,...,\gamma_p)$                           \\ \hline
{\gls{ows}}    & $\left\{
    \begin{tabular}{l}
        $ (\gamma_1,...,\gamma_i,...,\gamma_p)\to(\gamma_1,...,\gamma_i\pm\pi,...,\gamma_p)$ \\
        $(\beta_1,...,\beta_i,...,\beta_p)\to(\beta_1,...,-\beta_i,...,-\beta_p)$
    \end{tabular}
    \right.$         \\ \hline
 \multirow{2}{*}{Integer weights} &   $(\gamma_1,...,\gamma_i,...,\gamma_p)\to(\gamma_1,...,\gamma_i\pm2\pi,...,\gamma_p)$                                                    \\ \cline{2-2}
                                 & $(\beta_1,...,\beta_i,...,\beta_p)\to(\beta_1,...,\beta_i\pm\pi/2,...,\beta_p)$  \\ \hline
  {Real weights}
  & $\left\{
    \begin{tabular}{l}
        $(\gamma_1,...,\gamma_i,...,\gamma_p)\to-(\gamma_1,...,\gamma_i,...,\gamma_p)$ \\
        $(\beta_1,...,\beta_i,...,\beta_p)\to-(\beta_1,...,\beta_i,...,\beta_p)$
    \end{tabular}
    \right.$                       
\end{tabular}
\end{ruledtabular}
\end{table}

\paragraph*{Graphs where the sum of weights connected to each vertex is even.}
The cost in Eq.~(\ref{eq:Cp}) is invariant under parameter changes $\gamma_i\to\gamma_i\pm\pi$ for any layer $i$ in \gls{qaoa}, for graphs where the sum of the weights connected to each vertex is even, i.e., $\Sigma_j w_{jk}\in2\mathbb{Z},\forall k\in V$ (see proof in Appendix~\ref{app:symmetry}). 
This even weight sum (EWS) symmetry captures the previously known result for unweighted graphs where all vertices have even degree~\cite{lotshawEmpiricalPerformanceBounds2021}. More examples that exhibit this parameter pattern include graphs with only even weights and graphs where all vertices have even degrees with only odd weights, for instance, $w_{ij}\in\{-1,1\}$.

\paragraph*{Graphs where the sum of weights connected to each vertex is odd.}
The cost in \gls{qaoa} is invariant under the simultaneous parameter change $\gamma_i\to\gamma_i\pm\pi$ and $\beta_j\to -\beta_j$ in all layers $j\geq i$, for graphs where the sum of the weights connected to each vertex is odd, i.e., $\Sigma_jw_{jk}\in\mathbb{Z}\setminus2\mathbb{Z},\forall k\in V$ (see proof in Appendix~\ref{app:symmetry}).
This odd weight sum (OWS) symmetry contains the case of unweighted graphs when all vertices have odd degree~\cite{lotshawEmpiricalPerformanceBounds2021}. Moreover, we also include graphs where each vertex has an odd degree and all weights are odd, for example, $w_{ij}\in\{-1,1\}$.

As previously mentioned, these symmetries relate to patterns in the optimization landscape and help identify sets of optimal parameters that lead to the same \gls{qaoa} performance for a given MaxCut problem instance.
The optimal parameters are known analytically for unweighted $d$-regular triangle-free graphs at depth $p=1$~\cite{wangQuantumApproximateOptimization2018}, and empirically for the unweighted 3-regular tree subgraph up to $p=11$~\cite{wurtzFixedangleConjecturesQuantum2021}. Similar empirical results are known for unweighted regular graphs of degree $d\leq11$ at depth $p=2$~\cite{wurtzFixedangleConjecturesQuantum2021}, and exhaustive sets of small graphs with $n\leq9$ vertices at depths $p\leq3$~\cite{lotshawEmpiricalPerformanceBounds2021}. In Fig.~\ref{fig:angles_symmetry}, we show the parameter space domains that contain the optimal sets for unweighted regular graphs with even and odd degrees. We observe that, in both cases, the domains $\mathcal{U}_1=[0,\pi/2)\times[0,\pi/4)$ and $\mathcal{U}_2=[-\pi/2,0)\times[-\pi/4,0)$ contain a set of optimal parameters.

\paragraph*{Symmetry patterns at higher depths.}
At higher algorithm depths $p$, graphs that obey either the \gls{ews} symmetry or \gls{ows} symmetry will have $2^{p+1}$ areas containing an optimal set of parameters within $\mathcal{A}^p$. 
As shown in Table~\ref{tab:angle_symmetry_general}, the cost function will be invariant under the $m$-th algorithm layer parameter changes $\gamma_m \to \gamma_m\pm\pi$ for instances with \gls{ews} symmetry, and $\gamma_m \to \gamma_m\pm\pi$ and $\beta_i\to -\beta_i$ for $i\geq m$ for instances with \gls{ows} symmetry. These symmetries provide two optimal parameter choices at each layer, either the initial set or the symmetric one, which results in $2^p$ possible combinations of optimal parameters. 
Lastly, the cost also remains invariant under the ``time-reversal symmetry'', which changes the sign of all parameters in the algorithm, thus doubling the number of optimal choices. In total, there are $2^{p+1}$ sets of optimal parameters within $\mathcal{A}^p$ providing the same \gls{qaoa} performance. We could visualize them as different domains in the energy landscape, as shown in Fig.~\ref{fig:angles_symmetry} for $p=1$. 

In summary, we have identified parameter symmetries in \gls{qaoa} for weighted MaxCut instances that explain optimal parameter patterns from numerical simulations. 
Given the proven symmetries, for a given instance, all sets of optimal parameters can be retrieved knowing a single set, provided there is a single optimum in the domain $[0,\pi/2)\times[-\pi/4,\pi/4)$.
The choice of parameters in the algorithm determines the quantum circuit. Therefore, a higher number of optimal parameter sets results in more choices of quantum circuits and gates, potentially improving the hardware implementation of the algorithm.

\section{Transferability of optimal parameters}
\label{sec:transferability}

In this section, we analyze the performance of \gls{qaoa} when transferring the different sets of optimal parameters identified in Sec.~\ref{sec:SymmetriesQAOA} from one problem instance to another. 
To that end, we define a figure of merit to quantify how transferable the parameters are. We study this transferability property both analytically and numerically for several families of graphs, with different degree and weight distributions.

Previous research showed the viability of transferring optimal parameters of \gls{qaoa} for MaxCut between unweighted $3$-regular graphs of different sizes~\cite{brandaoFixedControlParameters2018, wurtzFixedangleConjecturesQuantum2021}, from unweighted to weighted graphs~\cite{sureshbabuParameterSettingQuantum2024, shaydulinParameterTransferQuantum2023}, between graphs with different degrees~\cite{galdaSimilarityBasedParameterTransferability2023,galdaTransferabilityOptimalQAOA2021} and between different \glspl{cop}~\cite{montanez-barreraTransferLearningOptimal2024,montanez-barreraUniversalQAOAProtocol2024}.
The transferability of optimal parameters can be explained by the decomposition of a given graph instance into local subgraphs at shallow \gls{qaoa} depths. Following a light cone argument, one can show that the algorithm sees a sum of local subgraphs instead of the whole graph. Thus, the \gls{qaoa} cost function only depends on these subgraphs, and the algorithm will perform similarly for two different problem instances with similar local subgraphs. For example, the cost function of Eq.~(\ref{eq:Cp}) for all unweighted $3$-regular graphs for \gls{qaoa} with depth $p=1$ is a sum of the cost of three subgraphs, times their respective multiplicity~\cite{farhiQuantumApproximateOptimization2014}. Such locality arguments for shallow algorithm depths also explain concentration properties of \gls{qaoa}, and have been linked to performance limitations for certain problems~\cite{farhiQuantumApproximateOptimization2020,anshuConcentrationBoundsQuantum2023,cerezoDoesProvableAbsence2023,bravyiObstaclesVariationalQuantum2020}.

The number of possible subgraphs for a $d$-regular graph grows exponentially with the algorithm depth $p$, if \gls{qaoa} does not see the whole graph, i.e., if $p\sim\mathcal{O}(\log n)$ with $n$ the number of vertices.
However, despite this exponential growth, in a sufficiently large random $d$-regular graph, almost all subgraphs will be tree subgraphs~\cite{lykovSamplingFrequencyThresholds2023, farhiQuantumApproximateOptimization2020,cainQAOAGetsStuck2023}, i.e., connected subgraphs graphs with no cycles. Consequently, the optimal \gls{qaoa} parameters for a $d$-regular graph will be close to the optimal parameters of the $d$-regular tree subgraph, which explains the successful parameter transferability between them at shallow depths. 
However, this argument fails to explain whether we can reuse optimal parameters between graphs with different degrees (or average degrees). 
Particularly, prior research suggests that, on average, we can successfully transfer any optimal parameters between regular graphs with the same parity (both have either odd or even degrees). However, for graphs with opposite parity, we can only reuse a subset of the optimal parameters~\cite{galdaSimilarityBasedParameterTransferability2023,galdaTransferabilityOptimalQAOA2021}. 

\paragraph*{Transferability error.}
To measure the success of reusing parameters of \gls{qaoa}, we define the transferability error $\transfer{\recep}{\donor}$ as
\begin{equation}
\label{eq:transferdef}
    \transfer{\recep}{\donor} = r_\recep(\gammapar_\recep^*,\betapar_\recep^*)-r_\recep(\gammapar_\donor^*,\betapar_\donor^*).
\end{equation} 
The error $\transfer{\recep}{\donor}$ is the difference between the approximation ratios $r_\recep$ for a receiver problem instance $\recep$---defined in Eq.~(\ref{eq:approxratioq})---when using its optimal parameters $(\gammapar^*_\recep,\betapar_\recep^*)$, and when using parameters that are optimal to a donor instance $\donor$, $(\gammapar^*_\donor,\betapar_\donor^*)$. 
A low error indicates a successful transfer.

\paragraph*{Analytic expression of the transferability error for simple cases at $p=1$.}
We begin our study of the parameters' transferability with the simplest example of unweighted $d$-regular graphs, for which we can rely on previously known analytical results.  
Given the locality of \gls{qaoa} at shallow depths, we choose the $d'$-regular tree subgraph---the most common subgraph---as the donor instance. Indeed, using the optimal parameters for the $3$-regular tree subgraph is known to give performance guarantees for arbitrary $3$-regular graphs at $p=1$~\cite{farhiQuantumApproximateOptimization2014} and $p=2$~\cite{wurtzFixedangleConjecturesQuantum2021}. 
Furthermore, the cost value for MaxCut with triangle-free graphs at depth $p=1$ has a known closed form~\cite{wangQuantumApproximateOptimization2018}. Using the closed form, we can derive an analytical expression for the transferability error between a $d'$-regular tree subgraph and a $d$-regular graph with girth $g>3$, i.e., triangle-free regular graphs. Thus, following our definition of Eq.~(\ref{eq:transferdef}), and identifying the receiver ($R$) and donor ($D$) graphs by their degree $d$ and $d'$, we have for depth $p=1$
\begin{equation}\label{eq:deltaapprox}
    \transfer{d}{d'}^{g>3}=\frac{1}{2k\sqrt{d}}\biggl[\left(\frac{d-1}{d}\right)^{\tfrac{d-1}{2}}-\sqrt{\frac{d}{d'}}\left(\frac{d'-1}{d'}\right)^{\tfrac{d-1}{2}}\biggr],
\end{equation}
with $k=C_\text{max}/|E|$, the fraction of edges cut in the maximum cut. In general, $k\in(1/2,1]$ and $k=1$ for bipartite graphs, and Eq.~(\ref{eq:deltaapprox}) is valid for sufficiently large graphs~\cite{farhiQuantumApproximateOptimization2014}, that is, graphs of order larger than $n\sim (d-1)^{2p}$  (see Appendix~\ref{app:Deltad} for a detailed derivation). 

\begin{figure}
    \centering
    \includegraphics[width=0.9\linewidth]{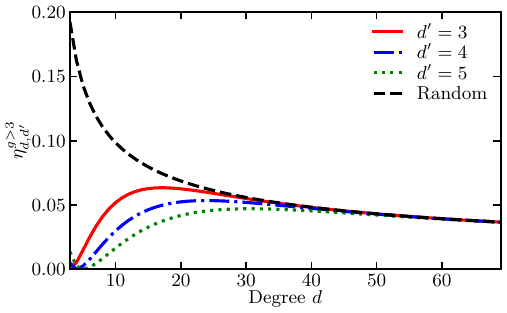}
    \caption{Transferability error $\transfer{d}{d'}^{g>3}$ from Eq.~(\ref{eq:deltaapprox}) for $d$-regular bipartite graphs ($k=1$) with a donor tree subgraph with degrees $d'=3$ (red solid line), $d'=4$ (blue dash-dotted line) and $d'=5$ (green dotted line). The black dashed line represents the transferability error when \gls{qaoa} with donated parameters is equivalent to random guessing.}
    \label{fig:delta_bipartitep1}
\end{figure}

The function in Eq.~(\ref{eq:deltaapprox}) allows us to calculate the transferability error for regular graphs up to an arbitrary degree. We visualize this expression in Fig.~\ref{fig:delta_bipartitep1} for bipartite receiver graphs with degrees up to $d=70$ and tree donor subgraphs of degree $d'\in\{3,4,5\}$. 
We also include a reference line where \gls{qaoa} with donated parameters is equivalent to random guessing, that is, when \gls{qaoa} gives an approximation ratio of $\frac{1}{2}$, or equivalently $\transfer{R}{D}=r_R(\gammapar_R^*,\betapar_R^*)-\frac{1}{2}$. Note that if we shift this line by $\frac{1}{2}$, it corresponds to the approximation ratio of \gls{qaoa} with optimal parameters, $r_R(\gammapar_R^*,\betapar_R^*)$, for increasing graph degrees. The transferability error $\transfer{d}{d'}^{g>3}$ is minimized for donor and receiver graphs with the same degree $d=d'$, in agreement with previous results~\cite{brandaoFixedControlParameters2018,galdaSimilarityBasedParameterTransferability2023}.
We observe that as the difference between the receiver and donor degrees $d$ and $d'$ grows, \gls{qaoa} with the parameters transferred from the tree subgraphs performs as random guessing with an approximation ratio of $r=1/2$~\cite{diestelGraphTheory2006}. Therefore, while the error tends to $0$ as $d\to\infty$, transferring parameters becomes pointless in the large $d$ limit. Moreover, when the error eventually reaches $0$ for large $d$, it also indicates that the approximation ratio of \gls{qaoa} with depth $p=1$ for any optimal parameters is $r\approx 1/2$~\cite{farhiQuantumApproximateOptimization2020}.

\subsection{Numerical analysis}
\label{subsec:transferabilitysim}
Here, we simulate and study the transferability of optimal \gls{qaoa} parameters, characterized by Eq.~(\ref{eq:transferdef}), for different families of graphs. 

\begin{figure}
    \centering
    \includegraphics[width=0.9\linewidth]{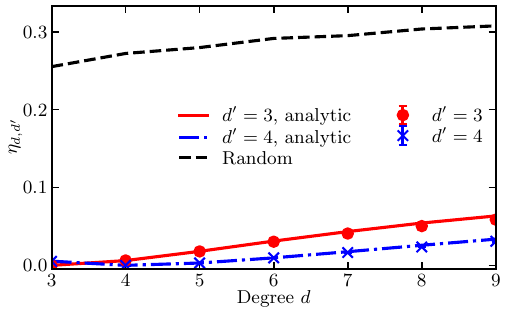}
    \caption{Average transferability error $\transfer{d}{d'}$ of Eq.~(\ref{eq:transferdef}) at depth $p=1$ for $d$-regular receiver graphs with $n=120$ vertices and donor tree subgraphs with degrees $d'=3$ (red circle) and $d'=4$ (blue cross). The error bars overlap with the symbols and denote the 25th and 75th percentiles.
    Comparison with the analytic results $\transfer{d}{3}^{g>3}$ (red solid line) and $\transfer{d}{4}^{g>3}$ (blue dashed-dotted line) of Eq.~(\ref{eq:deltaapprox}).
    The black dashed line indicates the average transferability error when QAOA with donated parameters performs as random guessing. }
    \label{fig:delta_randomp1}
\end{figure}

\paragraph*{Transferability from the $d'$-regular tree subgraph to $d$-regular unweighted graphs at $p=1$.}
This case is approximated by the analytic expression in Eq.~(\ref{eq:deltaapprox}), visualized in Fig.~\ref{fig:delta_bipartitep1}.
Here, we consider ten instances of order $n=120$ of receiver regular graphs for each degree ranging from $d=3$ to $d=9$. We use the optimal parameters for the triangle-free graphs, equivalent to the optimal parameters of the tree subgraphs, $\gamma_{d'}^*=\arctan(1/\sqrt{d'-1})$ and $\beta_{d'}^* =\pi/8$~\cite{wangQuantumApproximateOptimization2018}.
In Fig.~\ref{fig:delta_randomp1}, we observe that Eq.~(\ref{eq:deltaapprox}) reproduces the numerical simulations of Eq.~(\ref{eq:transferdef}), and that the transferability error increases when the difference between the degrees of the donor and receiver graphs grows.

We note that, in this case, the success of the transfer does not depend on the relative parity between the degrees $d$ and $d'$. Indeed, the set of optimal parameters from the donor graph lies in the domain $\mathcal{U}_1$ (see Fig.~\ref{fig:angles_symmetry}) and, as discussed in Sec.~\ref{sec:SymmetriesQAOA}, $\mathcal{U}_1$ contains optimal parameters for regular graphs with both odd and even degree.
Previous works also suggest that only parameters from such shared domains, $\mathcal{U}_1$ and $\mathcal{U}_2$, are the ones transferable between unweighted regular graphs with degrees of different parity~\cite{galdaSimilarityBasedParameterTransferability2023}.
However, considering the problem symmetries, if we instead have access to an optimal set of parameters from a different domain, it can be easily translated into either $\mathcal{U}_1$ or $\mathcal{U}_2$ following the rules of Table~\ref{tab:angle_symmetry_general}. In view of this, knowing the problem symmetries and the shared domains between instances, we can transfer any optimal parameter between any $d$-regular graph independently of the graphs' degrees. 

\paragraph*{Transferability from the $d'$-regular tree subgraph to random graphs at $p=1$.}
In this case, we use again the unweighted $d'$-regular tree subgraph as a donor graph, for general random receiver graphs.
Unlike regular graphs, random graphs do not have a fixed degree and, at low depths, \gls{qaoa} sees a local subgraph with a varying degree around each vertex. 
The random graph models will have different distributions of degrees. 
Here, we study three models of random graphs: the Erd\H{o}s-Rényi (ER) model~\cite{erdosRandomGraphs2022}, the Barabási-Albert (BA) model~\cite{albertStatisticalMechanicsComplex2002}, and the Watts-Strogatz (WS) model~\cite{wattsCollectiveDynamicsSmallworld1998}. 
While these models can be tuned to have the same average degree, they exhibit distinct properties that make them suitable for describing different systems.
The ER model generates random graphs where edges appear independently with equal probability, resulting in a binomial degree distribution. In contrast, the BA model generates scale-free networks characterized by large hubs, where a few vertices have many connections and most have few, following a power-law degree distribution. 
The WS model produces graphs with local clusters of connected vertices by rewiring edges in a ring lattice with fixed probability. 
Both BA and WS graphs are representative of real-world network structures.

\begin{figure}
    \centering
    \includegraphics[width = 0.9\linewidth]{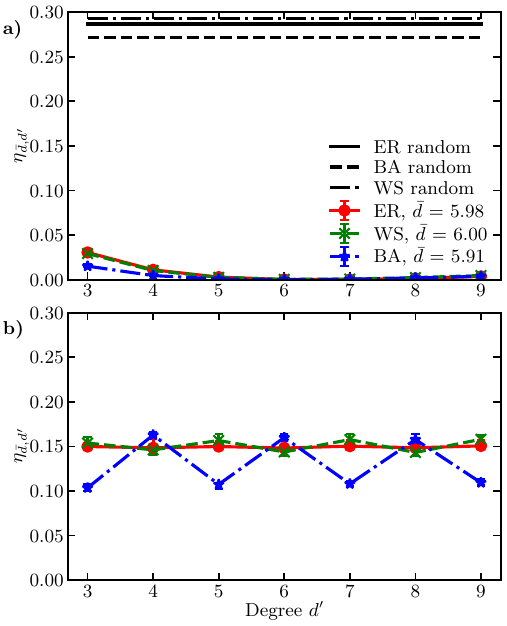}
    \caption{Average transferability error $\transfer{\bar{d}}{d'}$ of Eq.~(\ref{eq:transferdef}) at depth $p=1$ for random receiver graphs with $n=200$ vertices and average degree $\bar{d}\approx6$, sampled from the ER model (red solid line), the WS model (green dashed line), and the BA model (blue dash-dotted line). We consider donor regular tree subgraphs with degrees $3 \leq d'\leq 9$ and their optimal parameters from \textbf{a)} $\mathcal{U}_1$, and \textbf{b)} neither $\mathcal{U}_1$ nor $\mathcal{U}_2$. The error bars overlap with the symbols and denote the 25th and 75th percentiles. The black lines in \textbf{a)} represent the average transferability error when \gls{qaoa} with donated parameters is equivalent to random guessing for ER (solid line), WS (dashed line), and BA (dash-dotted line).
   }
    \label{fig:random_graphs}
\end{figure}

We use ten instances of each type of random receiver graph with order $n=200$, and the same average degree $\bar{d}\approx6$. The degree of the donor tree subgraph takes the values $d'=\{3,...,9\}$.
In contrast to the previous case, these random unweighted graphs have neither the OWS nor the EWS symmetry described in Sec.~\ref{sec:SymmetriesQAOA}, and we find that a transfer is only successful if the donor parameters are from the domains $\mathcal{U}_1$ or $\mathcal{U}_2$.
As we observe in Fig.~\ref{fig:random_graphs}, the transferability error is low for parameters from $\mathcal{U}_1$ and increases for parameters from neither $\mathcal{U}_1$ nor $\mathcal{U}_2$. Given the degree distribution of the models, the random graphs have both odd and even degree subgraphs. As a result, even the parameters optimal to the $6$-regular tree subgraph perform poorly if they are not included in $\mathcal{U}_1$ or $\mathcal{U}_2$.
We can compare this result to Ref.~\cite{galdaSimilarityBasedParameterTransferability2023}, where they fix the percentage of even-regular subgraphs in $20$-node graphs and find different \gls{qaoa} performances for parameters transferred from each domain. 
As in the previous case, the transferability error is minimized when the donor and receiver graphs have similar degrees $d'=\bar{d}$, that is, $\transfer{\bar{d}}{6}\approx0$ when the donor is a $6$-regular tree in our analysis. Furthermore, for the random graph models we studied, the impact of the variance of the degree distribution compared to the average degree is negligible. 
Additionally, the relative parity between the degree of the donor subgraph and the average degree only affects the transferability for the BA graphs, when using optimal parameters outside $\mathcal{U}_1$ or $\mathcal{U}_2$.

\paragraph*{Transferability from the $d'$-regular tree subgraph to $d$-regular graphs with integer weights at p=1.}

We study the transferability of parameters from unweighted tree donor subgraphs with degrees $d'=3$ and $d'=4$ to $d$-regular graphs with uniformly distributed integer weights $w_{ij}\in\{-1,1\}$, and degrees $d$ varying from $3$ to $9$. To this end, we use ten instances of $20$-node receiver graphs for each degree $d$. We choose the optimal parameter set of the donor graphs from the domain $\mathcal{U}_1$.
Following the discussion in Sec.~\ref{sec:SymmetriesQAOA}, we know that the receiver regular graphs with this weight distribution follow the \gls{ews} or the \gls{ows} symmetry for even or odd degrees, respectively.

\begin{figure}
    \centering
    \includegraphics[width=0.9\linewidth]{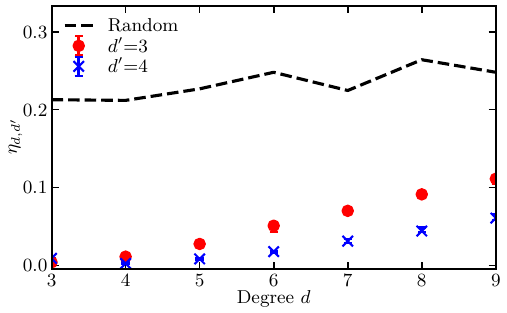}
\caption{Average transferability error $\transfer{d}{d'}$ of Eq.~(\ref{eq:transferdef}) at $p=1$ for $d$-regular receiver graphs with $n=20$ vertices, and uniformly distributed weights $w_{ij}\in\{-1,1\}$. The donor graphs are unweighted tree subgraphs with degrees $d'=3$ (red circle), and $d'=4$ (blue cross). The error bars overlap with the symbols and denote the 25th and 75th percentiles. The black dashed line shows the average transferability error when \gls{qaoa} with transferred parameters behaves as random guessing.}
    \label{fig:weighted_transfer}
\end{figure}

In Fig.~\ref{fig:weighted_transfer}, we observe a low transferability error, especially when $d\approx d'$, confirming that the donor parameters are transferable to the integer weighted graphs.
The small size of the simulated graphs ($n=20$) increases the numerical results' variability and moves us out of the large graph regime with triangle-free graphs~\cite{farhiQuantumApproximateOptimization2020}, especially for the higher degrees. Without this large graph approximation, the tree subgraph is a less suitable donor.
As in all previous examples, the error increases as the difference between $d$ and $d'$ increases and, in this case, it grows faster than for the unweighted $d$-regular graphs. 
The smaller sizes of the receiver graphs can contribute to this faster deterioration of the transferability. 
Additionally, we confirm with an initial numerical exploration that the transferability error increases when using parameters outside $\mathcal{U}_1$ and $\mathcal{U}_2$.
Since the symmetries that apply are the same as for the unweighted $d$-regular graphs, we can follow a similar argumentation as in the unweighted graphs case. 
Namely, the universally transferable domains for the weighted graphs are those universal to the unweighted $d$-regular graphs, i.e., $\mathcal{U}_1$ and $\mathcal{U}_2$. 
However, we know $d$-regular graphs with these integer weights satisfy either the \gls{ows} or the \gls{ews} symmetry, and using the rules in Table~\ref{tab:angle_symmetry_general} we can shift a given set of parameters so that the transferability becomes independent of the relative parity between $d$ and $d'$.

\paragraph*{Transferability at higher algorithm depths.}
As we show in Sec.~\ref{sec:SymmetriesQAOA}, at depth $p=2$, \gls{qaoa} for MaxCut on $d$-regular graphs has eight sets of optimal parameters in $\mathcal{A}^2$, instead of the four sets at $p=1$. 
We present these parameters in Appendix~\ref{app:optparamp2} (see Table~\ref{tab:p2parameters}) for the $3$-regular tree subgraph.
Here, we study the performance of \gls{qaoa} for unweighted regular graphs of order $n=120$ and degrees $d\in\{3,4,5\}$, considering ten graphs per degree, and using---without shifting to a common domain---the eight sets of optimal parameters of the $3$-regular tree donor subgraph.

\begin{figure}
    \centering
    \includegraphics[width=0.9\linewidth]{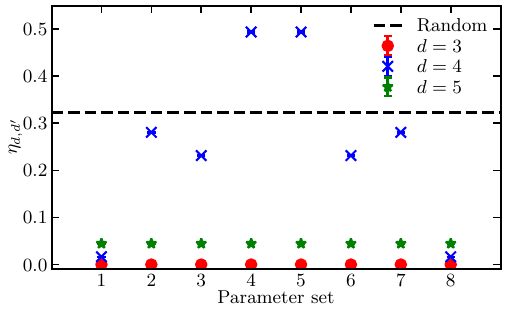}
    \caption{
    Average transferability error $\transfer{d}{d'}$ of Eq.~(\ref{eq:transferdef}) at depth $p=2$ for unweighted regular receiver graphs of degrees $d=3$ (red circles), $d=4$ (blue crosses), and $d=5$ (green stars), with an unweighted tree donor subgraph of degree $d'=3$. The error bars overlap with the symbols and denote the 25th and 75th percentiles. We use the eight optimal sets of parameters of the donor graph (see Table~\ref{tab:p2parameters}), numbered from one to eight, as shown on the $x$-axis. The sets $1$ and $8$ lie in $\mathcal{U}_1$ and $\mathcal{U}_2$, respectively. The black dashed line shows the average transferability error when QAOA with transferred parameters performs as random guessing.}
    \label{fig:p2}
\end{figure}

In Fig.~\ref{fig:p2}, we observe an analogous behavior to the case of $p=1$. That is, only parameters from the higher dimensional domains $\mathcal{U}_1^{\, 2}=[0,\pi/2)^2\times[0,\pi/4)^2$ and $\mathcal{U}_2^{\, 2}=[-\pi/2,0)^2\times[-\pi/4,0)^2$ can be transferred (without using symmetry rules) independently of the relative parity between the donor and receiver. 
When transferring these parameters to a receiver graph of degree $d=3$ or $d=5$, i.e., with the same parity, the error remains low for all eight sets of parameters. Conversely, when transferring to $4$-regular graphs, the error is significantly higher for parameters from neither $\mathcal{U}_1^{\, 2}$ nor $\mathcal{U}_2^{\, 2}$. We notice that these six sets of parameters could be translated into $\mathcal{U}_1^{\, 2}$ or $\mathcal{U}_2^{\, 2}$ using the symmetry relations of the cost function.
Unlike the case of tree subgraphs at $p=1$, there is no analytically known set of optimal parameters in $\mathcal{U}_1$ for a general $d$-regular tree at $p=2$. However, as mentioned in Sec.~\ref{sec:SymmetriesQAOA}, previous studies show numerical evidence for optimal parameters in $\mathcal{U}_1^{\, 2}$ for $d\leq 11$ and $p=2$~\cite{wurtzFixedangleConjecturesQuantum2021}. When restricting to the $3$-regular tree subgraph, the same authors provide numerically optimized values in $\mathcal{U}_1^{\, p}$ at even higher depths, $p\leq11$~\cite{wurtzFixedangleConjecturesQuantum2021}.
Moreover, other works that consider interpolation and pseudo-adiabatic paths for finding parameters at higher depths use the initial optimal parameters in $\mathcal{U}_1$ and find that $(\gammapar,\betapar)\in\mathcal{U}_1^{\, p}$~\cite{zhouQuantumApproximateOptimization2020, leeDepthProgressiveInitializationStrategy2023}. 
These adiabatic interpolation techniques fail when considering initial optimal parameters to $3$-regular graphs outside $\mathcal{U}_1$ and $\mathcal{U}_2$, $(\gamma^*,\beta^*)\in[\pi/2,\pi)\times[0,\pi/4)$~\cite{leeDepthProgressiveInitializationStrategy2023}.

Summarizing, prior work confirms the existence of optimal parameters in $\mathcal{U}_1^{\, p}$ for the tree subgraph up to $p=11$. However, more work is necessary to conclude whether these parameters are universally transferable at depths $p>2$ for regular graphs of any degree, either odd or even. For depths $p>11$, it remains an open question whether optimal parameters (transferrable or not) exist in $\mathcal{U}_1^{\, p}$. For instance, if we consider a linear ramp interpolating the optimal values of the parameters, we can expect that their values will eventually lie outside $\mathcal{U}_1^{\, p}$.
Nevertheless, any optimal parameter at those depths will follow the symmetry rules discussed in Sec.~\ref{sec:SymmetriesQAOA}.

\section{Parameter symmetries in other variational quantum algorithms}
\label{sec:extension}
The quantum circuit of many hybrid classical-quantum algorithms can be built through several iterations of similar quantum operations, e.g., any layer of QAOA contains a parametrized mixing unitary operation and a parametrized cost unitary operation. Such structured quantum circuit ansätze also appear in realizations of \glspl{vqe} for finding ground states of quantum many-body problems, as the \gls{hva}. Compared to the hardware-efficient ansatz, the problem-specific \gls{hva} circuit provides an easier optimization, reducing the challenges of barren plateaus while maintaining a high expressivity~\cite{PRXQuantum.1.020319}.
While in this paper our focus mainly lies on the \gls{qaoa} for different MaxCut instances and the transferability of its optimal parameters between various ensembles of graphs, we discuss here how these results can be extended to other \glspl{cop} and \glspl{vqa}. 
The \gls{ews} and \gls{ows} symmetries shown in Sec.~\ref{sec:SymmetriesQAOA} for the MaxCut problem also apply to all Ising Hamiltonians of the form $\hat{H}_C = \sum_{i<j} w_{ij} \pauli{i}{z}\pauli{j}{z}/2+\sum_i h_i\pauli{z}{i}/2$ (see detailed derivation in Appendix~\ref{app:otherCOPs}). The condition on the coefficients $w_{ij}$ and $h_i$ change to $\sum_j w_{jk}+h_k\in 2\mathbb{Z},\forall k \in V$ and $\sum_j w_{jk}+h_k\in \mathbb{Z}\setminus2\mathbb{Z},\forall k \in V$ for \gls{ews} and \gls{ows}, respectively. This extension covers a wide range of problems, including Karp's 21 NP-complete problems~\cite{lucasIsingFormulationsMany2014} and finding ground states of spin glass models.

The quantum unitary operations of the \gls{hva} use terms of many-body Hamiltonians, analogously to the cost function in QAOA and adiabatic quantum computation. 
Therefore, we can identify similar symmetries in these \glspl{vqa}. 
A notable example is the \gls{hva} for the \gls{tfim}, which coincides with the \gls{qaoa} for solving MaxCut in the ``ring of disagrees", a $2$-regular unweighted ring graph~\cite{farhiQuantumApproximateOptimization2014}.
Therefore, the \gls{hva} for the \gls{tfim} exhibits \gls{ews}, and the parameters can be optimized over a smaller domain. 
In Appendix~\ref{app:otherVQAs}, we show that \gls{ews} is valid for ansätze constructed with operators of the form 
\begin{equation}
\label{eq:Uaa}
    \hat{U}_{\alpha\alpha}(\gamma)=\prod_{(i,j)\in E}\exp(-i\frac{w_{ij}}{2}\gamma \pauli{\alpha}{i}\pauli{\alpha}{j}),
\end{equation} 
where $\alpha \in\{x,y,z\}$, given that the weights $w_{ij}$ fulfil the criteria of \gls{ews} described in Sec.~\ref{sec:SymmetriesQAOA}. In other words, any \gls{vqe} ansätz with these terms and conditions follows the \gls{ews}.

We extend our results to other \gls{hva} based on different Hamiltonian terms. In particular, we consider the general case of a Heisenberg model on high-dimensional spaces, where the connectivity between the spins is defined by the set of edges $E$ of a given graph. Moreover, we allow the coupling constants between the spins to vary from site to site, thus introducing disorder. The Hamiltonian reads
\begin{equation}
\label{eq:Heis_disorder}
\hat{H}= \frac{1}{2}\sum_{(i,j)\in E} \left( w_{ij}^{x} \pauli{x}{i}\pauli{x}{j} + w_{ij}^{y} \pauli{y}{i}\pauli{y}{j} + w_{ij}^{z} \pauli{z}{i}\pauli{z}{j}\right), 
\end{equation}
with $w_{ij}^{x}$, $w_{ij}^{y}$, and $w_{ij}^{z}$ the coupling strengths associated with the $x$-, $y$-, and $z$-components of the spin interaction between the sites $i$ and $j$, given by the edge $(i,j)$.
This general case reduces to well-known models for studying magnetic systems. For instance, we can consider a quantum spin chain with real-valued coupling constants, such that the Hamiltonian becomes
\begin{equation}
\label{eq:1dHeisenberg}
    \hat{H}_{\rm{XYZ}}= \sum_{i=1}^{n} \left(J_x \pauli{x}{i}\pauli{x}{i+1} + J_y \pauli{y}{i}\pauli{y}{i+1} + J_z \pauli{z}{i}\pauli{z}{i+1} \right),
\end{equation}
in which we assume periodic boundary conditions ($1\equiv n+1$). This one-dimensional Heisenberg XYZ model corresponds to a particular case of Eq.~(\ref{eq:Heis_disorder}) for a $2$-regular ring graph with $n$ vertices defining the connectivity of the interactions, and uniform coupling constants $w_{ij}^{\alpha}/2=J_{\alpha}$, with $\alpha\in\{x,y,z\}$.
When $J_x=J_y$, the model is called the one-dimensional Heisenberg XXZ model, studied in the context of \gls{hva}~\cite{PRXQuantum.1.020319}. 
Without loss of generality, we can take $J_x=J_y=1$ and $J_z=\Delta$, with $\Delta$ controlling the spin anisotropy. The critical point $\Delta = 1$ becomes harder to approximate classically~\cite{PRXQuantum.1.020319}.  
Interestingly, optimal \gls{hva} parameters have been successfully transferred from small to large instances of the Heisenberg XXZ model for parameters in the domain $\mathcal{U}_1^{\, p}$~\cite{PhysRevA.106.L060401}.
We derive symmetry results for all these models in the case of integer-valued coupling constants $w_{ij}^{\alpha}$. In Appendix~\ref{app:otherVQAs}, we base our derivations on the fact that the \gls{hva} for the general Hamiltonian of Eq.~(\ref{eq:Heis_disorder}) comprises products of operators given by Eq.~(\ref{eq:Uaa}). 

Finally, we also consider different versions of the \gls{qaoa}, namely the \gls{rqaoa}~\cite{bravyiObstaclesVariationalQuantum2020} and the \gls{ma-qaoa}~\cite{herrmanMultiangleQuantumApproximate2021}. The \gls{rqaoa} reduces the original optimization problem size by imposing constraints obtained from recursive applications of \gls{qaoa}. With each algorithm iteration, the weights and regularity of the problem graph change, altering the cost Hamiltonian for the \gls{qaoa} subroutine. 
As we show in Appendix~\ref{app:otherVQAs}, \gls{rqaoa} preserves the \gls{ews} of the original cost Hamiltonian throughout every iteration---each reduced cost Hamiltonian adheres to the \gls{ews}---but breaks the \gls{ows} after the first iteration.
This behavior implies that all optimal parameters are directly transferable if the original graph obeys \gls{ews}. Otherwise, for initial problem graphs with \gls{ows}, we can see from our discussion on transferability to random model graphs in Sec.~\ref{sec:transferability}, that only optimal parameters from the domains $\mathcal{U}_1$ or $\mathcal{U}_2$ can be reused between iterations.
Another version of \gls{qaoa} is the \gls{ma-qaoa}, which uses a higher number of parameters at shallow depths to create a more expressive ansatz.
Unlike standard \gls{qaoa}, which requires \gls{ews} or \gls{ows} for the entire graph, with \gls{ma-qaoa} we can focus on the weight sums of individual vertices to analyze the transferability of the gate parameters related to Hamiltonian terms of specific vertices and edges. 
Both \gls{rqaoa} and \gls{ma-qaoa} improve the performance of the standard \gls{qaoa} at shallow depths, albeit with increased demands on parameter optimization~\cite{herrmanMultiangleQuantumApproximate2021, bae2024improvedrecursiveqaoasolving, bravyiObstaclesVariationalQuantum2020,Gaidai_2024}. Therefore, insights from transferability analyses to improve these optimization loops are especially valuable.

\section{Conclusions and outlook}
\label{sec:outlook}
Optimizing the variational circuits remains a big obstacle for realizable and successful \glspl{vqa}. 
One useful simplification in the optimization task is limiting the search space using symmetries in the energy landscape. 
Additionally, the optimal parameters concentrate on typical values for different instances of certain problems. 
Thus, we can use these typical values as pretrained parameters and transfer them between instances, either to be used directly, or as initialization points for a warm-start optimization. In this article, we extend the known energy landscape symmetries of MaxCut and show how to leverage them to transfer optimal \gls{qaoa} parameters.
While previous studies focus on unweighted graphs, we find that for all graphs where the sum of the weights connected to each vertex is even, the landscape follows the \gls{ews} symmetry (see Table~\ref{tab:angle_symmetry_general}). Analogously, the landscape obeys the \gls{ows} symmetry for all graphs where the sum of the weights connected to each vertex is odd.
Among the examples following these symmetries, we find regular graphs with weights $w_{ij}\in\{-1,1\}$.

Regarding the optimal parameters for $p=1$, it is known that unweighted triangle-free regular graphs have an optimal set in the domain $\mathcal{U}_1=[0,\pi/2)\times[0,\pi/4)$~\cite{wangQuantumApproximateOptimization2018}, and from the symmetries, other optimal sets in the green solid areas of Fig.~\ref{fig:angles_symmetry}, depending on the parity of the degree.
Here, we analyze the transferability of these optimal parameters, i.e., those of a $d'$-regular tree subgraph, to triangle-free $d$-regular graphs, $d$-regular graphs, different kinds of random graphs, and $d$-regular graphs with integer weights. Our work shows that, in all cases, the graphs are viable recipients for the transferred parameters.

To measure the success of transferring parameters, we define a transferability error in Eq.~(\ref{eq:transferdef}), $\transfer{\recep}{\donor}$, as a function of the approximation ratios of \gls{qaoa} with the optimal parameters of the receiver ($\recep$) and donor ($\donor$) graphs. Using an analytically known set of optimal parameters to the $d'$-regular tree subgraphs, we derive a closed-form expression for the error when transferring to triangle-free $d$-regular graphs at $p=1$. The expression allows us to analyze the performance of transferability for large graphs with high $d$ without needing to simulate large \gls{qaoa} circuits. Furthermore, this approximated expression reproduces the numerical simulations of low-girth graphs. 
The transferability error is minimized and close to zero if $d\approx d'$ for $d$-regular graphs with a $d'$-regular donor tree subgraph. We also find that the error goes to zero in the $d\to \infty$ limit. 
However, we can explain the latter result by \gls{qaoa} performing equally poorly for all parameters.

We observe that the success of a transfer is independent of the relative parity between the degree of the donor and receiver when the transferred set of parameters comes from $\mathcal{U}_1$ or $\mathcal{U}_2$ (see Fig.~\ref{fig:angles_symmetry}), as already reported in previous works~\cite{galdaSimilarityBasedParameterTransferability2023}. However, we provide the rules to translate any optimal set of parameters into these domains using the donor graph symmetry, thus making all sets of optimal parameters transferable in practice.

Similarly, our results with random receiver graphs show that a transfer is only successful if the donor parameters come from $\mathcal{U}_1$ or $\mathcal{U}_2$. 
In all the cases with random graphs---generated from the ER model, the BA model, and the WS model---the success depends on the closeness between the degree of the donor and the average degree of the receiver graph, rather than on the degree distribution of the model.
In this case, we find that a transfer is only successful if the donor parameters come from $\mathcal{U}_1$ or $\mathcal{U}_2$, independently of the relative parity between donor and receiver graphs. 
When considering weighted graphs obeying either the \gls{ews} or the \gls{ows} symmetry, the transferability error behaves similarly to the case of unweighted graphs. 
That is, the minimum error occurs when $d=d'$, and the domains $\mathcal{U}_1$ and $\mathcal{U}_2$ contain optimal parameters that work regardless of the relative parity between the degrees of the donor and receiver graphs. 

At the higher algorithm depth $p=2$, we also find successful transfers between any $d$-regular and $d'$-regular graph, if the parameters come from the higher dimensional domains $\mathcal{U}_1^{\, 2}=[0,\pi/4)^2\times[0,\pi/2)^2$ or $\mathcal{U}_2^{\, 2}=[-\pi/4,0)^2\times[-\pi/2,0)^2$. 
Similarly, in this case, we can translate any optimal set of parameters to these domains using symmetries. 
We notice that the number of domains that contain an optimal set of parameters within $\mathcal{A}^p=[-\pi,\pi)^p\times[-\pi/4,\pi/4)^p$ scale as $2^{p+1}$, and our numerical simulations suggest that the number of shared domains suitable for transferring parameters is two. However, more studies are needed to conclude this.

Finally, we discuss the applicability of our results to other \glspl{cop} and \glspl{vqa}. We find conditions such that adding a local field term to the cost Hamiltonian in \gls{qaoa} preserves the two symmetries \gls{ews} and \gls{ows}, thus including general \glspl{cop} described by Ising Hamiltonians~\cite{lucasIsingFormulationsMany2014}. Additionally, we show that the EWS is valid for all unitaries associated with Hamiltonians with $\pauli{x}{}\pauli{x}{}$, $\pauli{y}{}\pauli{y}{}$, or $\pauli{z}{}\pauli{z}{}$ interactions. Among these Hamiltonians, we consider the general case of the Heisenberg model on high-dimensional spaces with integer-valued couplings. For specific coupling values, this model reduces to the XYZ and XXZ models, previously studied with gls{hva}~\cite{PRXQuantum.1.020319}. We also examine two \gls{qaoa} variants, \gls{rqaoa} and \gls{ma-qaoa}, which improve the performance of the standard \gls{qaoa} but require more parameter optimizations. We find that the iterative process of \gls{rqaoa} only maintains the \gls{ews}, but breaks the \gls{ows} after the first iteration. Furthermore, we observe that the \gls{ma-qaoa} could benefit from specific transferability schemes based on the symmetries of individual subgraphs, instead of the entire graph.

Our insights on the conditions of transferred parameters and their relation to the problem symmetries set the stage for a more comprehensive study of symmetry-based optimization in quantum algorithms. Future work may seek to numerically investigate the benefits of reducing the landscape size with these symmetries and performing warm-start local optimization with transferred parameters. For instance, the next steps would involve characterizing the number of iterations in the classical optimization compared to the current standard approaches.

\begin{acknowledgments}

We acknowledge useful discussions with Zeidan Zeidan and Göran Johansson of \gls{qaoa} and our results. We acknowledge useful discussions with Werner Dobrautz and Erika Magnusson of the \glspl{vqa} used in chemistry. We acknowledge support from the Knut and Alice Wallenberg Foundation through the Wallenberg Centre for Quantum Technology (WACQT). L.G.-Á. further acknowledges funding under Horizon Europe programme HORIZON-CL4-2022-QUANTUM-01-SGA via the project 101113946 OpenSuperQPlus100.
\end{acknowledgments}

\appendix

\section{Simulation methods}
In this section, we detail the methods used to simulate the \gls{qaoa} and perform the classical optimization.
All graphs are generated using the Python package NetworkX~\cite{hagberg2008exploring}, and the figures are created with the Python package Matplotlib~\cite{hunterMatplotlib2DGraphics2007}.
We simulate MaxCut with $d$-regular graphs and random graphs on QAOA using the tensor-network-based package Qtensor~\cite{lykovQTensor2021, lykovPerformanceEvaluationAcceleration2021}. The optimization of the parameters was done with the Qtensor implementation of the Python package PyTorchs~\cite{paszkePyTorchImperativeStyle2019} optimizer RMSprop, with 50 optimization steps.

We simulated MaxCut on weighted $d$-regular graphs with QAOA using Qiskit~\cite{javadi-abhariQuantumComputingQiskit2024}, with 3000 shots per circuit evaluation. Here, the parameters' optimization was carried out with the Python package SciPys~\cite{2020SciPy-NMeth} and the optimizer 
differential\_evolution with a population size of 15, and a maximum of 1000 function calls. For all optimizations, we used the optimal parameters to the relevant $d$-regular tree subgraphs as initial guesses.

\section{Symmetries of optimal QAOA parameters on MaxCut}
\label{app:symmetry}

Here, we provide detailed proofs for the \gls{qaoa} parameter symmetries presented in Sec.~\ref{sec:SymmetriesQAOA}, extending the results presented in Ref.~\cite{lotshawEmpiricalPerformanceBounds2021} for unweighted graphs where all vertices have either odd or even degree. 
We identify two symmetries for regular graphs with integer weights that can be classified according to the parity of the weights' sum on each vertex. In the following subsections, we present the results for graphs in which the weights of the edges connected to each vertex sum to an even and odd quantity, respectively.

\subsection{Symmetries for graphs where the sum of weights connected to each vertex is even}
In this section, we prove the symmetry called \gls{ews} in Sec.~\ref{sec:SymmetriesQAOA}.
We consider a variational state constructed by a \gls{qaoa} circuit of depth $p$. We focus on the $m$-th iteration of the algorithm, in which the operation $\hat{U}_M(\beta_m)\hat{U}_C(\gamma_m)$ acts on the intermediate variational state 
\begin{equation}
\label{eq:m_iter}
    \ket{\Psi_{m-1}}= \prod_{\mathclap{j=m-1}}^{1} e^{-i\beta_j\mixerH}e^{-i\gamma_j\costH}\ket{+}.
\end{equation}

Let us now assume that all eigenvalues of the diagonal cost Hamiltonian $\costH$ are even, that is, for a computational basis $n$-qubit state $\ket{y}$---with $y=y_1y_2 \dots y_n$ using the binary representation---$\costH\ket{y}=2\ell_{y}\ket{y}$, for $\ell_{y} \in \mathbb{Z}$. 
In this case, the operations $\hat{U}_C(\gamma_m)$ and $\hat{U}_C(\gamma_m\pm \pi)$ are equivalent, as we can observe by their action on $\ket{\Psi_{m-1}} = \sum_{y=0}^{2^n -1} c_y \ket{y}$. Explicitly, 
\begin{align}
    e^{-i\gamma_m\costH}{\textstyle\sum_{y}} c_y \ket{y} & = {\textstyle\sum_{y}} c_y e^{-i\gamma_m 2\ell_y} \ket{y} \nonumber \\ 
    & = {\textstyle\sum_{y}} c_y e^{-i\gamma_m 2\ell_y} e^{\pm i2\ell_y \pi} \ket{y} \nonumber \\ 
    & = e^{-i(\gamma_m\pm\pi)\costH}{\textstyle\sum_{y}} c_y \ket{y},
\end{align}
where we have introduced the global phases $e^{\pm i2\ell_y \pi}=1$. That is, the output of the \gls{qaoa} is invariant under changes $\gamma_m\to\gamma_m\pm\pi$ if the costs associated with each computational state $\ket{y}$ are even.
For a graph instance of order $n$, the cost associated with the MaxCut problem introduced in Eq.~(\ref{eq:clasMaxCut}) evaluates to zero for the string of $n$ ones $x^1=1^{n}$, $C(x^1)=0$. Consequently, in the quantum formulation, $\costH$ will have the even eigenvalue zero for the product state $\ket{1}^{\otimes n}$, $\costH\ket{1}^{\otimes n}=0$. In the following, we consider an arbitrary string $x\in \{-1,1\}^{n}$ and show that changing the value of one bit changes its cost by an even amount. Therefore, since the cost of $x^1$ is even, all the costs are even in both the classical and quantum formulations, which in turn proves the symmetry.

The costs of Eq.~(\ref{eq:clasMaxCut}) for two strings $x$ and $x'$ that differ only by the $k$-th bit, such that $x'_k=-x_k$, are 
\begin{equation}
    C(x) = \frac{1}{2} \sum_{i\neq k}\sum_{j>i} w_{ij}(1-x_i x_j) + \frac{1}{2}\sum_{j} w_{kj} (1-x_k x_j), \nonumber
\end{equation}
and
\begin{equation}
    C(x') = \frac{1}{2} \sum_{i\neq k}\sum_{j>i} w_{ij}(1-x_i x_j) + \frac{1}{2}\sum_{j} w_{kj} (1-x'_k x_j), \nonumber
\end{equation}
respectively. Thus, the cost difference is 
\begin{equation}
\label{eq:Cdelta}
    C(x)-C(x')=\sum_j w_{kj} x'_k x_j .
\end{equation} 
Since $x'_k x_j=\pm 1$, this difference will be an even quantity if the sum of all weights connected to a vertex $k$ is even. This requirement is equivalent to saying that the number of edges with odd integer weights connected to a vertex $k$ is even. For the symmetry to hold for all bit changes, we require this criterion to hold for each vertex.

Particularly, any graph with only even integer weights will show the parameter symmetry $\gamma_m\to\gamma_m\pm \pi$, for any layer $m$, regardless of each vertex degree. Moreover, graphs with only odd weights where all vertices have even degree (thus including regular graphs of even degree) will share the same symmetry, with the specific examples of all odd weights $w_{ij}\in{\{-1,1\}}$, and unweighted graphs when all vertices have even degree. This last example is already proven in Ref.~\cite{lotshawEmpiricalPerformanceBounds2021}. In Sec.~\ref{subsec:transferabilitysim}, we study numerically the transferability of the parameters for regular graphs of even degree with weights $w_{ij}\in{\{-1,1\}}$.

%%%%%%%%%%%%%
\subsection{Symmetries for graphs where the sum of weights connected to each vertex is odd}
In this section, we prove the symmetry called \gls{ows} in Sec.~\ref{sec:SymmetriesQAOA}.
As in the previous case, we consider the general case of a \gls{qaoa} circuit of depth $p$ and how the change $\gamma_m \rightarrow \gamma_m\pm\pi$ in the $m$-th layer changes the final output state. Given the cost Hamiltonian $\costH$ of Eq.~(\ref{eq:HC}) for MaxCut on a weighted graph, the algorithm operation of layer $m$ is
\begin{align}
\label{eq:UCoddpi}
\nonumber
    \hat{U}_C(\gamma_m\pm\pi) = \prod_{\mathclap{(i,j)\in E}}\exp[-i\frac{w_{ij}}{2}(\gamma_m\pm\pi)(\mathbb{1}-\pauli{z}{i}\pauli{z}{j})]\\
   =\hat{U}_C(\gamma_m) \prod_{\mathclap{(i,j)\in E}} \exp(\mp i\frac{w_{ij}\pi}{2})\exp(\mp i\frac{w_{ij}\pi}{2}\pauli{z}{i}\pauli{z}{j}).
\end{align}
Here, we can ignore the global phase factors $\exp(\mp i w_{ij}\pi/2)$ and focus on the operational terms differing from $\hat{U}_C(\gamma_m)$, which can be rewritten as
\begin{equation*}
\label{eq:weightedcossin}
    \exp(\mp i\frac{w_{ij}\pi}{2}\pauli{z}{i}\pauli{z}{j})=\cos(\frac{w_{ij}\pi}{2})\mp i\sin(\frac{w_{ij}\pi}{2})\pauli{z}{i}\pauli{z}{j}.
\end{equation*}

We only consider graphs with integer weights $w_{ij}$, which can be either even or odd. If a given weight $w_{ij}$ is even, only the cosine differs from zero, which just adds a global phase. Otherwise, if a weight $w_{ij}$ is odd, only the sine function is different from zero, and the resulting operation of the $m$-th layer in \gls{qaoa} changes. For convenience, we rewrite Eq.~(\ref{eq:UCoddpi}) as
\begin{align}
\label{eq:UCoddpi_odd_even}
    \hat{U}_C(\gamma_m\pm\pi) =
   \hat{U}_C(\gamma_m) e^{i\kappa} e^{i\delta} \prod_{\mathclap{\substack{(i,j)\in E \\ w_{ij} \textrm{ odd}}}} \pauli{z}{i}\pauli{z}{j} ,
\end{align}
where the global phases encompass all the phase factors that do not change the final variational state. These phases include the contribution of all edges with even weights and the phase factors from the sine functions, 
\begin{align}\label{eq:eikappa}
    e^{i\kappa} &=\prod_{\mathclap{(i,j)\in E}}\exp(\mp i\frac{w_{ij}\pi}{2})\prod_{\mathclap{\substack{(k,\ell)\in E \\ w_{\ell k} \textrm{ even}}}}\cos(\frac{w_{\ell k}\pi}{2}),\\\label{eq:eidelta}
    e^{i\delta} &=  \prod_{\mathclap{\substack{(i,j)\in E \\ w_{ij} \textrm{ odd}}}}  i\sin(\mp\frac{w_{ij}\pi}{2}).
\end{align}
As we can observe from Eq.~(\ref{eq:UCoddpi_odd_even}), we recover the symmetry $\gamma_m \rightarrow \gamma_m\pm\pi$ for any graph with even integer weights, already proven in the previous section.

In the following, we address the case in which a graph can have both even and odd integer weights. We denote the set of edges connected to a vertex $k$ as $E_k$ and focus on that vertex and the weights of $E_k$. The contribution of these edges to the product in Eq.~(\ref{eq:UCoddpi_odd_even}) is
\begin{equation}
    \prod_{\mathclap{\substack{(i,k)\in E_k \\ w_{ik} \textrm{ odd}}}} \pauli{z}{i}\pauli{z}{k} =  \left(\pauli{z}{k}\right)^{\mu_k} \prod_{\mathclap{\substack{i|(i,k)\in E_k \\ w_{ik} \textrm{ odd}}}} \pauli{z}{i} ,
\end{equation} 
where $\mu_k$ is the number of weights of the set $E_k$ with an odd integer value. 
Now, we require that $\mu_k$ is either even or odd for all vertices in the graph, $k\in V$, that is, that every vertex has the same parity for the number of odd weighted edges connected to it. We can simplify Eq.~(\ref{eq:UCoddpi_odd_even}) as
\begin{align}
\label{eq:UCoddpi_mu}
    \hat{U}_C(\gamma_m\pm\pi) =
   \hat{U}_C(\gamma_m) e^{i\kappa} e^{i\delta} \prod_{\mathclap{k\in V}} \left(\pauli{z}{k}\right)^{\mu_k} ,
\end{align}
and we have two cases: if $\mu_k$ is even, then  $(\pauli{z}{k})^{\mu_k}=\mathbb{1}$; otherwise, if $\mu_k$ is odd, then $(\pauli{z}{k})^{\mu_k}=\pauli{z}{k}$. 
In the first case, with $\mu_k$ even for all $k\in V$, we recover again the symmetry rule $\gamma_m \rightarrow \gamma_m\pm\pi$ described in the previous section, since the sum of all weights connected to each vertex is even. 
We note that if the cost Hamiltonian was expressed with $\pauli{x}{}\pauli{x}{}$ or $\pauli{y}{}\pauli{y}{}$ products instead of $\pauli{z}{}\pauli{z}{}$, similar derivations would hold. We discuss this fact further in our analyses of other \glspl{vqa} in Appendix~\ref{app:otherVQAs}.

We explore now the case in which $\mu_k$ is odd for all vertices $k\in V$ and include the action of the entire $m$-th \gls{qaoa} layer, also with the mixer unitary $\hat{U}_M(\beta_m)$. In particular, we consider changing its parameter $\beta_m \rightarrow \beta'_m$ such that we find algorithm symmetries. The complete algorithm step is then given by
\begin{align}
\label{eq:UMUC_sym}
  \hat{U}_M(\beta'_m) \hat{U}_C(\gamma_m\pm\pi) =& \prod_{\mathclap{k\in V}}\left[\cos(\beta'_m) -i\sin(\beta'_m)\pauli{x}{k} \right]  \nonumber \\ &\times e^{i\kappa} e^{i\delta} \pauli{z}{k} \hat{U}_C(\gamma_m).
\end{align}
Using the relation $\pauli{z}{k}\pauli{x}{k}=-\pauli{x}{k}\pauli{z}{k}$, and $e^{i\theta}=e^{i\kappa} e^{i\delta}$, we can rewrite Eq.~(\ref{eq:UMUC_sym}) as
\begin{align}
\label{eq:UMUC_com}
  \hat{U}_M(\beta'_m) \hat{U}_C(\gamma_m\pm\pi) =  e^{i\theta} \prod_{\mathclap{k\in V}}\pauli{z}{k} \hat{U}_M(-\beta'_m) \hat{U}_C(\gamma_m),
\end{align}
which allows us now to study the action of these parameter changes in the final \gls{qaoa} expectation value of the objective Hamiltonian, $C(\gammapar,\betapar)$ in Eq.~(\ref{eq:Cp}). In particular, we recall the intermediate variational state $\ket{\Psi_{m-1}}$ of Eq.~(\ref{eq:m_iter}), and construct the final state for the modified parameters as
\begin{align}
\label{eq:final_state_1}
  \ket{\gammapar',\betapar'}=& \ e^{i\theta} \prod_{\mathclap{j=p}}^{\mathclap{m+1}} \hat{U}_M(\beta'_j) \hat{U}_C(\gamma_j) \prod_{\mathclap{k\in V}}\pauli{z}{k}  \nonumber \\
   &\times  \hat{U}_M(-\beta'_m) \hat{U}_C(\gamma_m) \ket{\Psi_{m-1}} ,
\end{align}
with $\gammapar'=(\gamma_1, \dots ,\gamma_{m-1},\gamma_m\pm\pi,\gamma_{m+1}, \dots, \gamma_p)$ and $\betapar'=(\beta_1, \dots ,\beta_{m-1},\beta'_m,\beta'_{m+1}, \dots, \beta'_p)$.
Using again the commutation relations of the Pauli operators, we can rewrite Eq.~(\ref{eq:final_state_1}) as
\begin{align}
\label{eq:final_state_2}
  \ket{\gammapar',\betapar'}= \ e^{i\theta} \prod_{\mathclap{k\in V}}\pauli{z}{k} \prod_{\mathclap{j=p}}^{\mathclap{m}} \hat{U}_M(-\beta'_j) \hat{U}_C(\gamma_j) \ket{\Psi_{m-1}} .
\end{align}   

The expectation value of the objective Hamiltonian for the new parameters, $C(\gammapar',\betapar')$, can be constructed from the probabilities of measuring each computational state $\ket{y}$, $P'(y) = \bra{\gammapar',\betapar'}\ket{y}\bra{y}\ket{\gammapar',\betapar'}$, with
\begin{align}
\label{eq:prob_y}
  P'(y) =& \bra{\Psi_{m-1}} \prod_{\mathclap{j=m}}^{\mathclap{p}} \hat{U}^{\dagger}_C(\gamma_j) \hat{U}^{\dagger}_M(-\beta'_j)   \prod_{\mathclap{k\in V}}\pauli{z}{k} \ket{y} \nonumber \\
  &\times \bra{y} \prod_{\mathclap{k\in V}}\pauli{z}{k} \prod_{\mathclap{j=p}}^{\mathclap{m}} \hat{U}_M(-\beta'_j) \hat{U}_C(\gamma_j) \ket{\Psi_{m-1}} .
\end{align}
Given that the $\pauli{z}{k}$ operators commute with $\ket{y}\bra{y}$, and $\prod_{k\in V} (\pauli{z}{k})^2 = \mathbb{1}$, the probabilities become
\begin{align}
  P'(y) = \bra{\gammapar,\betapar'} \ket{y} \bra{y} \ket{\gammapar,\betapar'},
\end{align}
with 
\begin{align*}
    \gammapar&=(\gamma_1, \dots, \gamma_p),  \\
    \betapar'&=(\beta_1, \dots ,\beta_{m-1},\beta'_m,\beta'_{m+1}, \dots, \beta'_p) .
\end{align*}
We observe that the \gls{qaoa} symmetry, for which $P'(y)=P(y)$, with $P(y)=|\braket{y}{\gammapar,\betapar}|^2$, occurs then for the simultaneous parameter changes
\begin{align}
\label{eq:sym_odd_weights}
  \gammapar &\rightarrow (\gamma_1, \dots, \gamma_m \pm \pi, \dots, \gamma_p), \nonumber \\
  \betapar &\rightarrow (\beta_1, \dots ,\beta_{m-1},-\beta_m,-\beta_{m+1}, \dots, -\beta_p).
\end{align}
Thus, graphs where every vertex has an odd number of odd weighted edges connected to it are symmetric under the parameter changes of Eq.~(\ref{eq:sym_odd_weights}).
In other words, graphs where the sum of weights connected to each vertex is odd will exhibit these parameter symmetries.

In particular, graphs with only odd weights where all the vertices have odd degree (thus also regular graphs of odd degree) will exhibit this symmetry, which includes the unweighted graphs studied in Ref.~\cite{lotshawEmpiricalPerformanceBounds2021}. 
We study numerically the transferability of the parameters for regular graphs of odd degree with weights $w_{ij}\in{\{-1,1\}}$ in Sec.~\ref{subsec:transferabilitysim}.

\section{Symmetries of optimal QAOA
parameters on general COPs}
\label{app:otherCOPs}
In this section, we extend the results of Appendix~\ref{app:symmetry} and find the conditions for which the symmetries \gls{ews} and \gls{ows}, introduced in Sec.~\ref{sec:SymmetriesQAOA}, still apply if the cost Hamiltonian includes local field terms. In other words, a general cost Hamiltonian in the form of an Ising Hamiltonian can result in the parameter landscape having \gls{ews} or \gls{ows}. Many \glspl{cop} can be mapped to an Ising Hamiltonian, including all of Karp's 21 NP-complete problems~\cite{lucasIsingFormulationsMany2014}.

A general Ising Hamiltonian with local field terms reads 
\begin{equation}
    \hat{H}_C=\frac{1}{2}\sum_{\mathclap{(i,j)\in E}}w_{ij}\pauli{z}{i}\pauli{z}{j}+\frac{1}{2}\sum_{i\in V}h_i\pauli{z}{i},
\end{equation} 
where $E$ is the set of edges in an interaction graph, and $V$ is the set of spins. The \gls{qaoa} unitary operation associated to this cost Hamiltonian is
\begin{align}
    \hat{U}_C(\gamma_m)&= \prod_{\mathclap{(i,j)\in E}}\exp(-i\frac{w_{ij}}{2}\gamma_m\pauli{z}{i}\pauli{z}{j}) \nonumber \\
    & \times\prod_{\mathclap{i\in V}}\exp(-i\frac{h_i}{2}\gamma_m\pauli{z}{i}).
\end{align}
We now consider the parameter shift $\gamma_m\pm\pi$, so that 
\begin{align}
\label{eq:Uc_localfield}
\nonumber
    \hat{U}_C(\gamma\pm\pi)&=\hat{U}_C(\gamma_m)\prod_{\mathclap{(i,j)\in E}}\exp(\mp i\frac{w_{ij}\pi}{2}\pauli{z}{i}\pauli{z}{j})\\
    &\times \prod_{\mathclap{i\in V}}\exp(\mp\frac{ih_i\pi}{2}\pauli{z}{i}).
\end{align} 
Analogously to Appendix~\ref{app:symmetry}, we assume that the coefficients $w_{ij}$ and $h_i$ are integers. Then, we can rewrite the terms of Eq.~(\ref{eq:Uc_localfield}) as
\begin{align}
    &\prod_{\mathclap{(i,j)\in E}}\exp(\mp i\frac{w_{ij}\pi}{2}\pauli{z}{i}\pauli{z}{j}) = e^{i\varepsilon}e^{i\delta}\prod_{k\in V}(\pauli{z}{k})^{\mu_k}, \\
    \label{eq:Hcosthipi}
    &\exp(\mp\frac{ih_i\pi}{2}\pauli{z}{i})=\cos\left(\frac{h_i\pi}{2}\right)\mp i\sin\left(\frac{h_i\pi}{2}\right)\pauli{z}{i},
\end{align} 
where 
\begin{equation}
    e^{i\varepsilon} = \prod_{\mathclap{\substack{(k,\ell)\in E \\ w_{\ell k} \textrm{ even}}}}\cos(\frac{w_{\ell k}\pi}{2}),
\end{equation} 
and $e^{i\delta}$ is given by Eq.~(\ref{eq:eidelta}).
For integer coefficients $h_i$, the right-hand side of Eq.~(\ref{eq:Hcosthipi}) is $\mathbb{1}$ if $h_i$ is even, and $\mp i\pauli{z}{i}$ if $h_i$ is odd. Thus, we get 
\begin{equation}
    \prod_{\mathclap{i\in V}}\exp(\mp\frac{ih_i\pi}{2}\pauli{z}{i})=e^{i\lambda}e^{i\nu}\prod_{\mathclap{k\in V}}(\pauli{z}{k})^{h_k},
\end{equation}
with the global phases
\begin{equation}
\nonumber
    e^{i\lambda} = \prod_{\mathclap{\substack{i\in V \\ h_i \textrm{ odd}}}} i\sin(\mp\frac{h_i\pi}{2}), \textrm{ and }
    e^{i\nu}=\prod_{\mathclap{\substack{i\in V \\ h_{i} \textrm{ odd}}}}\cos(\frac{h_i\pi}{2}).
\end{equation}

We can finally rewrite Eq.~(\ref{eq:Uc_localfield}) as
\begin{equation}
    \hat{U}_C(\gamma_m\pm\pi) = e^{i\alpha}\hat{U}_C(\gamma_m)\prod_{\mathclap{k\in V}}(\pauli{z}{k})^{\mu_k+h_k},
    \label{eq:Uc_final_local field}
\end{equation}
with the global phase $e^{i\alpha}=e^{i\varepsilon}e^{i\delta}e^{i\lambda}e^{i\nu}$.
Now, we can follow a similar argument as in Appendix~\ref{app:symmetry} considering the quantity $\mu_k+h_k$, instead of just $\mu_k$, which was the number of odd-valued weights connected to vertex $k$. 
We conclude that the \gls{ews} applies if $\mu_k+h_k$ is even-valued, and \gls{ows} applies if $\mu_k+h_k$ is odd-valued. 

%Similarly, we could define a quantity $\Sigma_k$, defined by the number of odd-valued coefficients connected to vertex $k$, i.e., the weights and local field. We can equivalently require that $\Sigma_k$ is odd for all vertices $k$ for a graph to obey \gls{ows}, and that $\Sigma_k$ is even for all vertices $k$ for a graph to obey \gls{ews}.

\section{Symmetries of optimal parameters in other VQAs}
\label{app:otherVQAs}
Here, we connect and extend the symmetries derived for the \gls{qaoa} for the MaxCut problem to the \gls{hva} for condensed matter problems, such as those described by Eq.~(\ref{eq:Heis_disorder}) in Sec.~\ref{sec:extension}.
Moreover, we include a detailed derivation for similar symmetry results in iterative quantum algorithms such as the \gls{rqaoa}.

\subsection{Symmetries for interaction graphs where the sum of weights
connected to each vertex is even}
\label{app:even symmetries}

We consider the unitary operator from Eq.~(\ref{eq:Uaa}), 
\begin{equation}\label{eq:UaaApp}
    \hat{U}_{\alpha\alpha}(\gamma) = \prod_{\mathclap{(i,j)\in E}}\exp(-i\frac{w_{ij}}{2}\pauli{\alpha}{i}\pauli{\alpha}{j}),
\end{equation} with $\alpha=\{x,y,z\}$, and $E$ the set of edges in an interaction graph. We assume that the weights satisfy $\sum_jw_{jk}=2\mathbb{Z},\forall k\in V$, where $V$ is the set of all spins with an interaction. Now, following the same line of arguments made in Appendix~\ref{app:symmetry}, we find that
\begin{align}\nonumber
    \hat{U}_{\alpha\alpha}(\gamma\pm\pi)=&\prod_{\mathclap{(i,j)\in E}}\exp(-i\frac{w_{ij}}{2}\pauli{\alpha}{i}\pauli{\alpha}{j})\\\nonumber
    =& \, \hat{U}_{\alpha\alpha}(\gamma)\prod_{\mathclap{(i,j)\in E}}\exp(\mp i \pi \frac{w_{ij}}{2}\pauli{\alpha}{i}\pauli{\alpha}{j})\\\nonumber
    =& \, \hat{U}_{\alpha\alpha}(\gamma)e^{i\kappa}e^{i\delta}\prod_{\mathclap{\substack{(i,j)\in E \\ w_{ij} \textrm{ odd}}}}\pauli{\alpha}{i}\pauli{\alpha}{j}\\
    =& \, \hat{U}_{\alpha\alpha}(\gamma)e^{i\kappa}e^{i\delta}\prod_{\mathclap{k\in V}}(\pauli{\alpha}{k})^{\mu_k}.
\end{align} The global phases $e^{i\kappa}$ and $e^{i\delta}$ are given by Eq.~(\ref{eq:eikappa}) and~(\ref{eq:eidelta}) respectivly, $\mu_k$ is the number of odd-valued weights of the set $E_k$. The initial assumption made on the weights imply that $\mu_k$ is even, so $(\pauli{\alpha}{k})^{\mu_k}=\mathbb{1}$, and we retrieve that, up to a global phase, 
\begin{equation}
    \hat{U}_{\alpha\alpha}(\gamma\pm\pi)=\hat{U}_{\alpha\alpha}(\gamma).
\end{equation}
It follows that ansätze made up of products of operators on the form of Eq.~\ref{eq:UaaApp} will have \gls{ews}, like the one-dimensional Heisenberg XYZ model.
\subsection{Symmetries for RQAOA}
\label{app:rqaoa}
In this section, we show that \gls{rqaoa} preserves \gls{ews}, but breaks \gls{ows}. The \gls{rqaoa} is a recursive algorithm in which, after each iteration, one of the vertices with the largest correlation magnitude is effectively removed from the graph.
Formally, if the maximal correlation is $|\mathcal{M}_{k\ell}|$, with $\mathcal{M}_{k\ell}=\bra{\gammapar,\betapar}\pauli{z}{k}\pauli{z}{\ell}\ket{\gammapar,\betapar}$, one of the vertices $k$ or $\ell$ is removed by imposing the constraint $\pauli{z}{\ell}=\text{sgn}(M_{k\ell})\pauli{z}{k}$. For our purpose, it is sufficient to consider $\pauli{z}{\ell}=\pm \pauli{z}{k}$. 

With \gls{rqaoa}, the dimension of the Hilbert space decreases every iteration. Thus, we can identify each distinct cost Hamiltonian of different iterations by the systems' size or number of qubits, $n$. Here, we label the initial Hamiltonian as $\hat{H}_n$, the Hamiltonian after the recursive step, $\hat{H}_{n-1}$, and so on. We are interested in how the sum of integer coefficients (weights and local field) connected to each vertex changes with each iteration. 
We recall from Sec.~\ref{sec:extension} that, for graphs with \gls{ews}, the sum of coefficients connected to any vertex $k$ is even, $\Sigma_jw_{jk}+h_k\in2\mathbb{Z},\forall k\in V$. In contrast, for graphs with \gls{ows}, the sum of coefficients connected to any vertex $k$ is odd, $\Sigma_jw_{jk}+h_k\in\mathbb{Z}\setminus2\mathbb{Z},\forall k\in V$.

Now, we consider an initial cost Hamiltonian $\hat{H}_n$ related to a graph that obeys either \gls{ews} or \gls{ows}. After the first iteration, we impose the constraint $\pauli{z}{\ell}=\pm\pauli{z}{k}$ such that the vertex $\ell$ is removed from the graph---or qubit $\ell$ is removed from the system. 
To analyze the symmetry of the new Hamiltonian $\hat{H}_{n-1}$, we first examine the terms of the initial cost Hamiltonian related to each vertex $i$, that is, the terms in which qubit $i$ participates. We denote such terms as $\hat{H}_n^i=\sum_jw_{ij}\pauli{z}{i}\pauli{z}{j}+h_i$, and the total cost Hamiltonian can be written then as $\hat{H}_n =\sum_i\hat{H}_n^i /2$.

For a given vertex $i$, we find that there are three possibilities: either $i$ share no edge with the removed vertex $\ell$, or $i$ share one edge with the removed vertex $\ell$, or $i=k$, meaning that it shares one edge with the removed vertex $\ell$, and that edge is removed by the constraint.
In the first case, if the vertex $i$ does not share an edge with vertex $\ell$, we observe that $\hat{H}_{n-1}^i=\hat{H}_n^i$.
Second, if the vertex $i$ shares an edge with vertex $\ell$ with the weight $w_{i\ell}$,  after the \gls{rqaoa} iteration, this vertex $i$ will share a new edge with vertex $k$ instead with weight $\pm w_{i\ell}$. Then, the initial Hamiltonian terms
\begin{equation}
\label{eq:Hnoneedge}
    \hat{H}_n^i=\sum_{j\neq \ell}w_{ij}\pauli{z}{i}\pauli{z}{j}+w_{i\ell}\pauli{z}{i}\pauli{z}{\ell}+h_i\pauli{z}{i},
\end{equation}
will change to
\begin{equation}
\label{eq:Hn1oneedge}
\hat{H}_{n-1}^i=\sum_{j\neq \ell}w_{ij}\pauli{z}{i}\pauli{z}{j}\pm w_{i\ell}\pauli{z}{i}\pauli{z}{k}+h_i\pauli{z}{i}.
\end{equation} 
Let us consider the difference in the sum of coefficients between Eq.~(\ref{eq:Hnoneedge}) and~(\ref{eq:Hn1oneedge}). The weights of the edges not connected to vertex $\ell$ and the local field $h_i$ remain unchanged, so the difference is given by $w_{i\ell}\pm w_{i\ell}$. This quantity is always even, as it is either equal to zero, or to $2w_{i\ell}$. Thus, the parity of the sum of coefficients around a vertex $i$ in this second case does not change by the constraint imposed in the recursive algorithm.
Lastly, we consider the vertices directly subject to the constraint, $\ell$ and $k$. The contribution to the initial cost Hamiltonian related to vertex $k$ is 
\begin{equation}
\label{eq:Hk}
\hat{H}_n^k=\sum_{j\neq \ell}w_{jk}\pauli{z}{j}\pauli{z}{k}+w_{\ell k}\pauli{z}{\ell}\pauli{z}{k}+h_k\pauli{z}{k},
\end{equation} 
while for vertex $\ell$ we have
\begin{equation}
\label{eq:Hell}
\hat{H}_n^\ell=\sum_{j\neq k}w_{j\ell}\pauli{z}{j}\pauli{z}{\ell}+w_{\ell k}\pauli{z}{k}\pauli{z}{\ell}+h_\ell\pauli{z}{\ell}.
\end{equation} 
As the graph follow either the \gls{ews} or the \gls{ows}, the parity of the sums of coefficients in Eqs.~(\ref{eq:Hk}) and (\ref{eq:Hell}) is the same. After the first \gls{rqaoa} iteration, we get
\begin{align}
\label{eq:Hn1lk}\nonumber
    \hat{H}_{n-1}^k&=\sum_{j\neq \ell}w_{jk}\pauli{z}{j}\pauli{z}{k}\pm\sum_{j\neq k}w_{j\ell}\pauli{z}{j}\pauli{z}{k}+(w_{\ell k}\pm w_{\ell k})\\
    &+(h_k\pm h_\ell)\pauli{z}{k},
\end{align} 
and $\hat{H}_{n-1}^\ell = 0$. 
The term $(w_{\ell k}\pm w_{\ell k})$ in Eq.~(\ref{eq:Hn1lk}) is a constant energy shift, and not a weight on an edge.

We focus now on the sum of coefficients connected to vertex $k$ in the Hamiltonian $\hat{H}_{n-1}^k$ of Eq.~(\ref{eq:Hn1lk}), denoted as $s_{n-1}(k)$, which is 
\begin{align}\label{eq:weightell}\nonumber
    s_{n-1}(k)=&\sum_{j\neq \ell}w_{jk}\pm\sum_{j\neq k}w_{j\ell}+(h_k\pm h_\ell)
    \\\nonumber=&\sum_{j\neq \ell}w_{jk}+w_{\ell k}+h_k\pm\sum_{j\neq k}w_{j\ell}\pm w_{\ell k}\pm h_\ell
    \\
    -&w_{\ell k}\mp w_{\ell k},
\end{align} where we have added and subtracted $w_{\ell k}$ twice.
The first three terms of the second line in Eq.~(\ref{eq:weightell}) are equal to $s_n(k)$, the sum of coefficients that connected to vertex $k$ in $\hat{H}_n^k$. Similarly, the next three terms correspond to $\pm s_n(\ell)$. 
We get 
\begin{equation}\label{eq:sn1k}
    s_{n-1}(k)= s_n(k)\pm s_n(\ell) -w_{\ell k}\mp w_{\ell k}.
\end{equation}
As the graph in $\hat{H}_n$ is assumed to be either \gls{ews} or \gls{ows}, $s_n(k)$ and $s_n(\ell)$ have the same parity, so the sum, or difference, of these quantities must be even. The quantity $-w_{\ell k}\mp w_{\ell k}$ is also even, making Eq.~(\ref{eq:sn1k}) even. 

To summarize, after the constraint is imposed, all vertices except one are related to coefficients that sum to a value with unchanged parity. The exception is vertex $k$, the one directly affected by the constraint in the previous example. Regardless of the symmetry of the initial Hamiltonian $\hat{H}_n$, the sum of coefficients associated with it after the iteration, $s_{n-1}(k)$, is even. As such, \gls{rqaoa} preserves \gls{ews}, but breaks \gls{ows}.

\section{Derivation of the analytic expression of the transferability error}
\label{app:Deltad}
In this section, we show a detailed derivation of the closed form expression for $\transfer{d}{d'}^{g>3}$ from Eq.~(\ref{eq:deltaapprox}). The expression is exact for unweighted triangle-free $d$-regular graphs, shown in Fig.~\ref{fig:delta_bipartitep1}, and approximate for unweighted low-girth $d$-regular graphs, shown in Fig.~\ref{fig:delta_randomp1}.

From Ref.~\cite{wangQuantumApproximateOptimization2018}, the objective function of an unweighted triangle-free graph is given by 
\begin{equation}
\label{eq:AppCd}
    C_{d}=\frac{|E|}{2}\left[1+\sin4\beta\sin\gamma\cos^{d-1}\gamma\right],
\end{equation} 
evaluated at $(\gamma_d^*,\beta_d^*)=(\arctan\frac{1}{\sqrt{d-1}},\frac{\pi}{8})$ yields \begin{equation}
\label{eq:AppCdmax}
    C_d^*=\frac{|E|}{2}\left[1+\frac{1}{\sqrt{d}}\left(\frac{d-1}{d}\right)^{\frac{d-1}{2}}\right].
\end{equation} 
The maximum number of edges that can be cut is $|E|$, and the minimum number of edges that can be cut is $\frac{|E|}{2}\left(1+\frac{1}{d}\right)-\frac{1}{4}$. For our purpose, it is sufficient to lower bound the cut to half the edges, so $\frac{C_{\text{max}}}{|E|}=k\in(0.5,1]$. The minimum cut value is $C_\text{min}=0$, so $r_d = \frac{C_d}{C_\text{max}}$. From the definition of $\transfer{d}{d'}$ from Eq.~(\ref{eq:transferdef}) we get \begin{equation}
   \transfer{d}{d'}=\frac{1}{k|E|}[C_d(\gamma_d^*,\beta_d^*)-C_d(\gamma_{d'}^*,\beta_{d'}^*)].
\end{equation}

Since we consider triangle-free receiver graphs, we replace $C_d(\gamma_d^*,\beta_d^*)$ with Eq.~(\ref{eq:AppCdmax}). The donor graph is the $d'$-regular tree subgraph so $C_d(\gamma_{d'}^*,\beta_{d'}^*)$ is Eq.~(\ref{eq:AppCd}) evaluated at $(\gamma_{d'}^*,\beta_{d'}^*)=(\arctan\frac{1}{\sqrt{d'-1}},\frac{\pi}{8})$. Then, we get
\begin{align}
     \transfer{d}{d'}^{g>3} =&\frac{1}{2k}\Biggr[\frac{1}{\sqrt{d}}\left(\frac{d-1}{d}\right)^{(d-1)/2} \nonumber \\
     &-\cos^{d-d'}\gamma_{d'}^*\frac{1}{\sqrt{d'}}\left(\frac{d'-1}{d'}\right)^{(d'-1)/2}\Biggr].
\end{align}
Using the trigonometry relation 
\begin{equation}
    \cos^n(\arctan x)=\left(\frac{1}{1+x^2}\right)^{\frac{n}{2}}, \nonumber
\end{equation}
we get the final expression 
\begin{align}
     \transfer{d}{d'}^{g>3}=&\frac{1}{2k\sqrt{d}}\Biggr[\left(\frac{d-1}{d}\right)^{(d-1)/2} \nonumber\\ 
     &-\sqrt{\frac{d}{d'}}\left(\frac{d'-1}{d'}\right)^{(d-1)/2}\Biggr].
\end{align} 
Note that the exponent is $\frac{d-1}{2}$ for both terms.

\section{Optimal parameters for the 3-regular tree subgraph at \texorpdfstring{$p=2$}{p=2}}
\label{app:optparamp2}
\begin{table}[ht]
\caption{\label{tab:p2parameters} Eight sets of optimal parameters to the $3$-regular tree subgraph at depth $p=2$ within $\mathcal{A}^2=[-\pi,\pi)^2\times[-\pi/4,\pi/4)^2$. Set 1 and 8 are in the domains $\mathcal{U}_1^{\, 2}$ and $\mathcal{U}_2^{\, 2}$, respectively.
}
\begin{ruledtabular}
\begin{tabular}{ccccc}
\textbf{} & $\gamma_1$ & $\beta_1$ & $\gamma_2$ & $\beta_2$  \\ 
Set 1    & $0.156\pi$   & $0.177\pi$ & $0.286\pi$ & $0.0933\pi$       
\\ \hline
Set 2    & $0.156\pi$   & $0.177\pi$ & $-0.714\pi$ & $-0.0933\pi$          \\ \hline
Set 3 & $0.844\pi$  & $0.177\pi$ & $0.714\pi$ & $-0.0933\pi$\\ \hline
Set 4 & $0.844\pi$ & $0.177\pi$ & $-0.286\pi$ & $0.0933\pi$\\ \hline 
Set 5 & $-0.844\pi$ & $-0.177\pi$ & $0.286\pi$ &  $-0.0933\pi$ \\ \hline
Set 6 & $-0.844\pi$ & $-0.177\pi$ &  $-0.714\pi$ & $0.0933\pi$\\ \hline
Set 7 & $-0.156\pi$ & $-0.177\pi$ & $0.714\pi$ & $0.0933\pi$ \\ \hline
Set 8 & $-0.156\pi$ & $-0.177\pi$ & $-0.286\pi$ & $-0.0933\pi$
\end{tabular}
\end{ruledtabular}
\end{table}
Here, we present the eight sets of optimal parameters to the $3$-regular tree subgraph at $p=2$ that we used as donors in Fig.~\ref{fig:p2} of Sec.~\ref{sec:transferability}. The sets are taken from Ref.~\cite{wurtzFixedangleConjecturesQuantum2021}, where the authors define the parameters $(\gammapar^*,\betapar^*)\in[0,2\pi)^2\times[0,\pi/2)^2$. However, to follow the convention used in this article, we have translated the parameters into $\mathcal{A}^p$ using the \gls{ows} symmetry. We provide the optimal parameter values in Table~\ref{tab:p2parameters}.

\bibliography{main.bbl}

\end{document}